\documentclass[lettersize,journal]{IEEEtran}
\usepackage{amsmath,amsfonts}
\usepackage{algorithmic}
\usepackage{algorithm}
\usepackage{array}
\usepackage[caption=false,font=normalsize,labelfont=sf,textfont=sf]{subfig}
\usepackage{textcomp}
\usepackage{stfloats}
\usepackage{url}
\usepackage{verbatim}
\usepackage{graphicx}
\usepackage{cite}
\usepackage{amssymb}
\usepackage{multirow}
\usepackage{makecell}
\usepackage{color}

\newcommand{\etal}{\textit{et al.}}
\newtheorem{Definition}{\textbf{{Definition}}}
\newtheorem{Proposition}{\textbf{{Proposition}}}
\begin{document}

\title{Defending Against Neural Network Model Inversion Attacks via Data Poisoning}
\author{Shuai~Zhou, 
        Dayong~Ye, 
        Tianqing~Zhu*,~\IEEEmembership{Member,~IEEE,}
        and~Wanlei~Zhou~\IEEEmembership{Senior Memeber,~IEEE}
\IEEEcompsocitemizethanks{\IEEEcompsocthanksitem Shuai Zhou, Tianqing Zhu, and Wanlei Zhou are with the Faculty of Data Science, City University of Macau,
Macau, China. *Tianqing Zhu is the corresponding author. 
E-mail: \{shuaizhou,~wlzhou,~tqzhu\}@cityu.edu.mo
\IEEEcompsocthanksitem Dayong Ye is with the Centre of Cyber Security and Privacy and the School of Computer Science, University of Technology Sydney, Ultimo, NSW 2007, Australia. E-mail: dayong.ye@uts.edu.au.}

\thanks{Manuscript received May 25, 2021.}}

\markboth{Journal of \LaTeX\ Class Files,~Vol.~14, No.~8, August~2021}%
{Shell \MakeLowercase{\textit{et al.}}: A Sample Article Using IEEEtran.cls for IEEE Journals}

\IEEEpubid{0000--0000/00\$00.00~\copyright~2021 IEEE}

\maketitle

\begin{abstract}
Model inversion attacks pose a significant privacy threat to machine learning models by reconstructing sensitive data from their outputs. While various defenses have been proposed to counteract these attacks, they often come at the cost of the classifier's utility, thus creating a challenging trade-off between privacy protection and model utility. Moreover, most existing defenses require retraining the classifier for enhanced robustness, which is impractical for large-scale, well-established models. This paper introduces a novel defense mechanism to better balance privacy and utility, particularly against adversaries who employ a machine learning model (i.e., inversion model) to reconstruct private data. Drawing inspiration from data poisoning attacks, which can compromise the performance of machine learning models, we propose a strategy that leverages data poisoning to contaminate the training data of inversion models, thereby preventing model inversion attacks.

Two defense methods are presented. The first, termed \textbf{l}abel-\textbf{p}reserving poisoning attacks for \textbf{a}ll output vectors (LPA), involves subtle perturbations to all output vectors while preserving their labels. Our findings demonstrate that these minor perturbations, introduced through a data poisoning approach, significantly increase the difficulty of data reconstruction without compromising the utility of the classifier. Subsequently, we introduce a second method, \textbf{l}abel-\textbf{f}lipping poisoning for \textbf{p}artial output vectors (LFP), which selectively perturbs a small subset of output vectors and alters their labels during the process. Empirical results indicate that LPA is notably effective, outperforming the current state-of-the-art defenses. Our data poisoning-based defense provides a new retraining-free defense paradigm that preserves the victim classifier's utility.
\end{abstract}


\section{Introduction}\label{sec:introduction}
\IEEEPARstart{M}{achine} learning revolutionized numerous domains. Given the surging demand for machine learning applications and the substantial computational resources required to power them, Machine-Learning-as-a-Service (MLaaS) has become a popular paradigm. Public users can access the trained model through APIs. However, the open accessibility of these models substantially amplifies concerns regarding data privacy~\cite{tanuwidjaja2020privacy,salem2020updates,MSurvey3,MSurvey5,S1,S2,S3}. In such contexts, model inversion attacks~\cite{Mahendran2015unbyinv,fredrikson2015mi,zhou2023boosting,zhu2022label,ye2024defending,zhou2024inversion} pose a particular threat, as they have the potential to reconstruct the original training data or sensitive features solely by utilizing the confidence vectors provided by the targeted models. 
\IEEEpubidadjcol
Numerous studies have explored mitigating the risks of model inversion attacks~\cite{xiao2020adversarial, MID2020,ye2022one,fredrikson2014privacy,fredrikson2015mi,salem2020updates,yang2020defending,wen2021defending}. Some propose retraining models with added regularization, like minimizing mutual information between inputs and outputs, to alleviate the risks of being inverted~\cite{xiao2020adversarial, MID2020}.  While effective, these methods require retraining and may be impractical for well-established models. Other approaches suggest modifying confidence vectors to increase model security~\cite{ye2022one,fredrikson2014privacy,fredrikson2015mi,yang2020defending,wen2021defending}. However, these often come at the cost of reduced model utility. MID~\cite{MID2020} provided theoretical analysis for the limitations of random noises on preserving model utility. Given these findings, we aim to develop defenses that enhance privacy protection without necessitating retraining and without diminishing the model's utility.

Data poisoning attacks could offer a novel defense against model inversion attacks. Data poisoning attacks typically aim to impair the training of a neural network by incorporating misleading data, thereby degrading the model's performance.  The susceptibility of deep learning models to such poisoned data is well-established. In the context of model inversion attacks, the inversion model is similarly vulnerable to such data manipulation. This vulnerability can be turned to our advantage to design countermeasures against model inversion attacks by introducing data poisoning attacks to corrupt the classifier's confidence vectors (i.e., training data of the inversion model). This method carefully targets only the information that could facilitate model inversion attacks, leaving other data untouched. This approach contrasts with differential privacy techniques that add random noise, which can indiscriminately compromise the relationship between the data and their confidence vectors. Our targeted data poisoning strategy, therefore, promises to preserve the model's utility more effectively than methods that indiscriminately obscure data with noise.

Along similar lines, adversarial examples have historically been employed to construct robust defenses against privacy invasions~\cite{DBLP:conf/ccs/JiaSBZG19, wen2021defending}. They usually interfere during a model's inference phase. However, in guarding against inversion attacks, defenders do not have access to the model's inference phase. Instead, defenders can only manipulate the training data of adversaries. Hence, our paper concentrates on poisoning attacks that target the training phase. Our strategy, grounded in data poisoning, hinders attackers' effective training of models, more directly addressing privacy issues. Understanding the gap between inversion and typical tasks like classification, we have adapted a popular data poisoning method for reconstruction tasks.  
Our contributions can be summarized as follows:
\begin{itemize}
    \item To the best of our knowledge, we are the first to introduce data poisoning attacks as a countermeasure against a model inversion attack. We explore how effective data poisoning is for creating defenses against model-based attacks. This strategy could also be applied to more attacks, such as membership inference attacks and attribute inference attacks.
    \item We adopt data poisoning to craft perturbations to replace the existing random noises, which provides a novel direction for improving the trade-off between utility and privacy in the context of noise-based defenses.
    \item We provide a flexible and practical defense for well-trained models that cannot be re-trained. Our experiments demonstrate that our defense can effectively boost privacy protection without re-training. 
\end{itemize}

\section{Related Work}
\subsection{Model Inversion Attacks}
Model inversion attacks aim to reconstruct examples (or some sensitive features) that share the same distribution as the original training data. Studies on model inversion can be roughly divided into two categories. The first uses gradient-based optimization techniques, often referred to as optimization-based algorithms~\cite{zhang2020secret,iccv/ChenKJQ21,struppek2022plug,Mirror2022}. Note, however, that most optimization-based model inversion attacks require a well-trained GAN (or training a GAN from scratch) and white-box access to the target model. This can be quite impractical and costly in real-world scenarios. The second category trains a separate model (i.e., the inverse of the victim model) to reconstruct the sensitive information. These are known a training-based attacks~\cite{DBLP:journals/corr/abs-1902-08552,iccv/ZhaoZXL21}. 

Yang \etal~\cite{DBLP:journals/corr/abs-1902-08552} proposed a strategy to train a neural network as the inverse of the target CNN. The inversion model takes the output of the target classifier as input and accurately recovers the original examples. Zhao \etal~\cite{iccv/ZhaoZXL21} proposed a method for a special scenario where the target task is different from the attack task. They believed that explanations (i.e., emotion prediction) improve the privacy risks for model inversion attacks. However, their methods are not suitable to the normal setting illustrated by Yang \etal~\cite{DBLP:journals/corr/abs-1902-08552}.
\subsection{Countermeasures against Model Inversion Attacks}
\subsubsection{Regularization-based defenses}
Xiao \etal~\cite{xiao2020adversarial} proposed an adversarial reconstruction learning framework to avoid the original example being reconstructed from the latent representations. They adopted an alternative strategy to train the target classifier and a substitute inversion model, through which the latent representation preserves the privacy of the original data. Wang \etal~\cite{MID2020} presented mutual information defense (MID) by introducing mutual information regularization in the training of the target classifier. This effectively reduces the dependency between the original examples and their predictions while retaining prediction power. Two additional metrics, F1-score and AUROC, are used to evaluate utility given a highly imbalanced dataset. A novel evaluation classifier then assesses the performance of the attack algorithms. Moreover, they also provided some theoretical analysis to formalize the model inversion attacks with two games and presented a theoretical explanation as to why defenses based on differential privacy (DP) do not confer reasonable utility.
\subsubsection{Fine-tuning predication defenses}
Given a well-trained model, some reactive defenses can be performed on the confidence vectors after the training phase. Naturally, Fredrikson \etal~\cite{fredrikson2014privacy} proposed to use DP as a way to protect the training data with a theoretical guarantee. Ye \etal~\cite{ye2022one} also introduced a differentially-private defense method to tackle membership inference attacks and model inversion attacks simultaneously. Their method also uses a DP mechanism - one that modifies the confidence vectors but the rank of scores in the normalized vector is retained, which means little classification accuracy is lost. Fredrikson \etal~\cite{fredrikson2015mi} suggested that reducing the precision of the target model can mitigate the threat of a model inversion attack. Salem \etal~\cite{salem2020updates} also proposed a possible defense by injecting noise into the confidence vectors to sanitize the posteriors. Yang \etal~\cite{yang2020defending} developed a purification framework to purify the confidence vectors. The framework reduces their dispersion and improves the robustness of the target model against inference attacks.

\begin{figure*}[!t]
    \centering
    \includegraphics[width=0.8\linewidth]{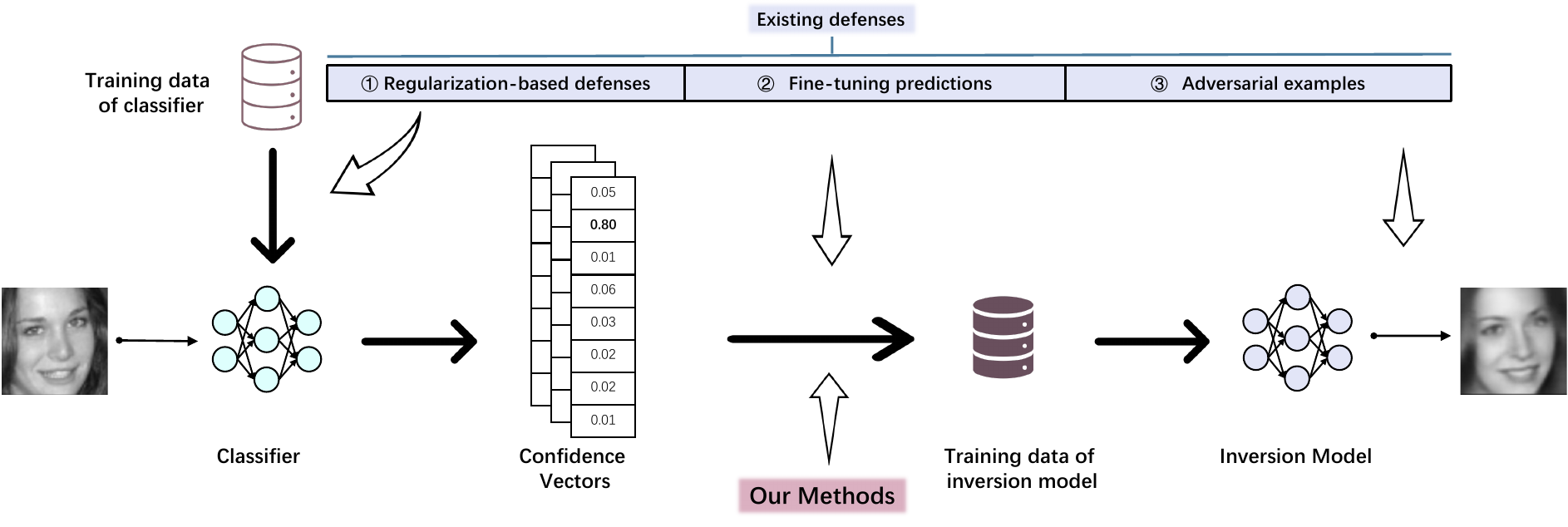}
    \caption{The pipeline of a model inversion attack. The test samples are fed into the classifier, and the classifier provides the confidence vectors for them. These predictions can be used by attackers to train an inversion model to reconstruct private images. Different defenses act at different phases. Regularization-based defenses tend to modify the training procedure of the classifier, while fine-tuning predictions and our methods manipulate the confidence vectors generated by the classifier. Adversarial examples are used to fool the inversion model in the reconstruction phase after it is trained.}
    \label{fig:pipeline}
\end{figure*}

One novel defensive technique adopts adversarial examples to fool the attacker's model~\cite{wen2021defending}. However, this technique fails to stop an effective attack model from being generated. Rather, it merely exploits the model's weaknesses to tempt it into making a mistake. As depicted in Fig.~\ref{fig:pipeline}, regularization-based defenses are implemented in the training of the classifier (i.e., the victim). Fine-tuning prediction defenses protect the well-trained classifier by manipulating the confidence vectors. However, defenses based on adversarial examples assume that an effective inversion model has been trained, and they come into play at the reconstruction phase of the inversion model.

\subsection{Adversarial Attacks}
Adversarial attacks have become a hot research topic recently due to the great vulnerabilities of machine learning models against adversarial examples. Adversarial attacks can be classified as evasion attacks or poisoning attacks based on the phase in which they are executed~\cite{chakraborty2021survey}. 
 
 \textcolor{black}{Data poisoning attacks are initially designed to compromise machine learning models, thereby degrading model performance and reliability~\cite{saha2020hidden,s5}. Recent studies on the intersection of privacy and security in machine learning can be categorized into two main directions. The first direction focuses on exploiting poisoning attacks to amplify privacy vulnerabilities. For instance, Truth Serum~\cite{TruthSerum} and Chen et al.~\cite{AmplifyMem} demonstrates how poisoning attacks can increase the success rate of membership inference attacks, while Chase et al. show how such attacks can enhance property information leakage~\cite{PropetryInf}. These studies primarily emphasize the malicious aspects of data poisoning in compromising privacy. 
 The second direction explores the benign applications of security attacks, particularly in defending against privacy attacks~\cite{jia2018attriguard,DBLP:conf/ccs/JiaSBZG19,wen2021defending,liutrap}. Most works in this category utilize adversarial attacks as defensive measures~\cite{jia2018attriguard,DBLP:conf/ccs/JiaSBZG19,wen2021defending}. However, a notable recent example, Trap-MID, proposes defending against model inversion attacks by embedding trapdoors through data poisoning attacks rather than adversarial attacks~\cite{liutrap}. While our method shares this defensive objective based on data poisoning, it differs fundamentally in implementation. Unlike Trap-MID requiring model training, our method implements defensive measures for trained models without altering the original training procedure. This key distinction makes our approach particularly valuable for real-world applications where model retraining is impractical or cost-prohibitive.}

\section{Preliminary}
\subsection{Classification Models}
The machine learning models that predict which class a target sample falls into are known as classifiers. In the training stage, a classifier is updated iteratively on a dataset with $m$ classes of data, $D_{tr}^{cla}$. The objective function $\mathcal{L}$ will lead to the optimal parameters $\theta$, which allow a trained classifier $\mathcal{T}$ to achieve satisfactory accuracy on unseen samples, $D_{test}^{cla}$. Thus, the well-trained classifier can provide an accurate prediction $c = \mathcal{T}(x) = (c_0, c_1,\cdots,c_{m-1})$ for a test sample $x$ in the inference phase. Conventionally, $c$ is a confidence vector with $m$ components and is the output of a softmax function. Accordingly, each component $c_i$ is an individual confidence score that satisfies $0\leq c_i\leq 1$ and $\sum_i c_i = 1$, which can be considered as the probability of class membership. The class with the maximal confidence score $c_i$ will become the predicted label.

\subsection{Data Poisoning Attacks}
Data poisoning attacks aim to render machine learning models inaccurate by polluting their training data. Here, the attacker injects bad or mislabeled examples into the training dataset. When the target model is trained on the corrupted data, some of the output it produces will be inaccurate~\cite{shafahi2018poison,schwarzschild2021just,shejwalkar2021manipulating}. In this paper, we focus on the data poisoning attacks targeting to cause substantial damage to the global performance of the target models~\cite{yang2017generative,munoz2017towards,muller2020data}. This category of attacks aims to maximize the average test loss, which can be formalized as a bi-level optimization with an inner learning problem as follows:
\begin{equation}
\centering
\begin{aligned}
    x^*_p &= \arg\max_{x_p} \mathcal{L}(\theta^*, D_{val})\\
    &s.t. ~\theta^* = \arg\min_\theta \mathcal{L}(\theta, D_{tr}\cup x_p)
\end{aligned}
\end{equation}
where $D_{tr}$ is the training data of the target model and $D_{val}$ is the validation dataset. $\theta$ represents the target model and $\theta^*$ is the optimal $\theta$ from the inner learning problem. These attacks that indiscriminately compromise global performance are more likely to be used to explore the vulnerability of attackers in model inversion tasks.

\section{Threat Model}
In our formulation, we use a setting similar to that in the work of Yang \etal~\cite{DBLP:journals/corr/abs-1902-08552}. Our focus is on training-based model inversion attacks due to their practicability in black-box scenarios. This type of attack aims to train an inversion model that takes the output of the victim model as its input and uses it to reconstruct the original input data, i.e., $D_{tr}^{cla}$~\cite{DBLP:journals/corr/abs-1902-08552}. The attackers intend to steal the users' semantic privacy, such as their face images, and subsequently their identities to circumvent any security systems. More specifically, considering an attacker who has only black-box access to the target model, the structure and parameters of the target model $\mathcal{T}$ are unknown to the attacker. To perform a model inversion attack against $\mathcal{T}$, the attacker can obtain a set of confidence vectors by sending query examples, $D_{tr}^{inv}$ to the target model. Based on the obtained data pairs, i.e., the confidence vectors and the query examples, the inversion model $\mathcal{I}$ can be trained successfully. The inversion model is then used to reconstruct the private data based on full confidence vectors~\cite{DBLP:journals/corr/abs-1902-08552}. 
Formally, the adversary aims to minimize the following loss:
\begin{equation}
\label{equ:obj}
    \mathcal{L}(\mathcal{I}, D_{tr}^{inv}) = \mathbb{E}_{x\in D_{tr}^{inv}} [\mathcal{R}(\mathcal{I}(\mathcal{T}(x)), x)]
\end{equation}
where $x$ is the query example (e.g., an image), and $\mathcal{T}(x)$ is its predicted confidence vector. $\mathcal{R}$ is used to measure the reconstruction error between the reconstructed image $\mathcal{I}(\mathcal{T}(x))$ and the original image $x$, e.g., the mean squared error (MSE). In addition to the reconstruction error $\mathcal{R}(\mathcal{I}(\mathcal{T}(x)), x)$, the second metric regarding the reconstruction quality is attack accuracy, which reflects the re-classification accuracy when the reconstructed image is fed into an evaluation model. Based on the first privacy metric (i.e., reconstruction error), we provide a formal definition of the success of model inversion attacks.

\begin{Definition} ($\gamma$-Inversion Attack)
Let $\mathcal{I}$ be the inversion model trained by an attacker, $D_{tr}^{cla}$ be the training data of the target model$\mathcal{T}$. A model inversion attack with $\mathcal{I}$ is called a successful $\gamma$-inversion attack against the target model $\mathcal{T}$ if the following property is held:
$$\mathbb{E}_{x\in D_{tr}^{cla}} [\mathcal{R}(\mathcal{I}(\mathcal{T}(x)), x)]\leq\gamma$$
where the threshold $\gamma$ is dataset-specific.
\end{Definition}

To eliminate the model inversion threats from malicious customers while preserving the utility of target models, the defender adds a post-processing unit $PU$ to a well-trained target model, without retraining the target model. The $PU$ can be implemented based on data poisoning attacks, as detailed in Section \ref{sec:method}, and works in the inference phase of the target models. 
Like DP-based defenses injecting noise~\cite{fredrikson2014privacy,ye2022one}, the $PU$ also modifies the confidence vectors for the query examples, regardless of whether an inverse adversary exists or not. Specifically, when the target model receives a query sample $x$, the output (i.e., the confidence vectors) $c = \mathcal{T}(x)$ are modified with a picking strategy $S$, which is also outlined Section \ref{sec:method}. As a result, the inference process of the target model under our defense is modified as follows:
\begin{equation}
\label{inference}
    c' = PU_\delta(\mathcal{T}(x)) =\left\{
    \begin{aligned}
    & \mathcal{T}(x) + \delta, && \text{~if~} S(x) = 1\\
    & \mathcal{T}(x), && \text{~if~} S(x) = 0
    \end{aligned}
    \right.
\end{equation}

While an attacker may conduct the attack and train the inversion model locally, they are required to construct the training data for the inversion model through interactions with the target model. Our defense strategies are designed to intervene during this critical interaction phase. With these defenses, we tampered with (some of) the predicted confidence vectors by adding perturbations, effectively poisoning the samples. These tampered outputs are then returned to users as the outcomes of their queries. Consequently, if an adversary utilizes this corrupted training data to construct an inversion model $\mathcal{I}$, that model would have compromised performance. 

In other words, the essence of our method lies in the contamination of this training data, which is introduced prior to the initiation of any attacks. This proactive approach ensures that the data used to train the inversion model is inherently compromised, regardless of how the training process is conducted. Consequently, even if the inversion model is trained locally, its quality and effectiveness are compromised due to the polluted training data.
\section{Methodology}
\label{sec:method}
\subsection{Overview}
\begin{figure*}[!t]
    \centering
    \includegraphics[width=0.80\textwidth]{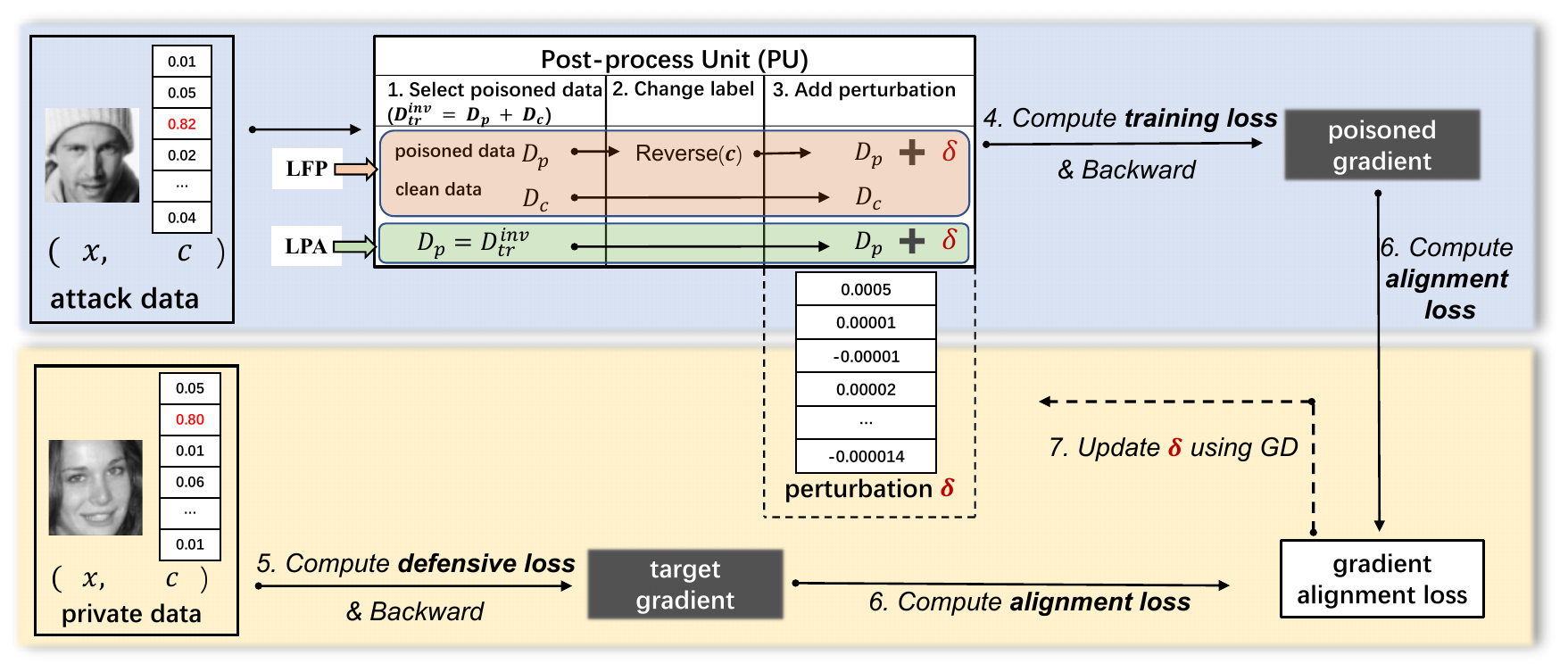}
    \caption{Overview of our unified framework of LPA and LFP. Through the Post-process Unit (i.e., step $1-3$), the attack data of the attacker are transformed into corrupted training data with the initial perturbation $\delta$. When the corrupted data are fed to the inversion model, the training loss can be computed on the perturbed data, and then the poisoned gradient can be further determined (step $4$). Likewise, the target gradient can be determined by calculating the defensive loss on private data (step $5$). To align these two types of gradients, the gradient alignment loss is computed, and its gradient w.r.t. the perturbation is utilized to update the perturbation based on the gradient descent (GD). The perturbation can be iteratively updated by repeatedly doing steps $1-7$. Notably, this Post-process Unit is a sort of unification. In particular, the PU of LFP differs from that of LPA. The former has all three steps (step $1-3$) while the latter contains only step $3$ and $D_c$ is $\emptyset$.}
    \label{fig:methods}
\end{figure*}

In this section, we explore how to perturb the confidence vectors based on data poisoning strategies. Two defense methods are discussed - one based on clean-label data poisoning (referred to as label-preserving poisoning for all vectors, abbreviated as LPA), and the other on non-clean-label data poisoning (referred to as label-flipping poisoning for partial vectors, abbreviated as LFP). Our LPA perturbs all the confidence vectors of the queried samples by adding minor noises. Namely, $S(x) =1$ for any $x$ in the Equation~\ref{inference}. 
By contrast, our second defense, LFP, leaves the vast majority of confidence vectors unchanged; only a small subset is selected to be processed. Compared to the LPA method, LFP replaces poisoned confidence vectors' corresponding images with new ones. The difference between these methods is depicted in Fig.~\ref{fig:methods}. In general, our strategy is manipulating the training data of the inversion model $\mathcal{I}$ and replacing the query results $c$ with $c'$ as mentioned in Equation~\ref{inference}. Our goal is to \textit{maximize} the following defensive loss. 
\begin{equation}
\label{equ:defensive}
\centering
\begin{aligned}
   \mathcal{L}_{def}(\mathcal{I}, D_{tr}^{cla}) &= - \mathcal{L}(\mathcal{I}, D_{tr}^{cla}) \\&= - \mathbb{E}_{x\in D_{tr}^{cla}} [\mathcal{R}(\mathcal{I}(c'), x)]\\
\end{aligned}
\end{equation}
where $c'$ is computed in Equation~\ref{inference}. And $\mathcal{I}$ is trained on the perturbed data pairs, $x \in D_{tr}^{inv}$ and corresponding $c'$.
\subsection{Method $1$: LPA defense}
Existing defenses typically improve the robustness against model inversion attacks by introducing random noises into the confidence vectors~\cite{ye2022one,fredrikson2015mi}. However, these random noises indiscriminately obfuscate the relation between the training data and their confidence vectors. Consequently, these countermeasures with random noises are more likely to undermine the utility of the target model. 

Our first method, LPA, seeks to compute perturbations $\delta$ to the output of the target model, which aims to degrade the performance of the inversion model $\mathcal{I}$, particularly the generalization to the private data $D_{tr}^{cla}$. This goal can be approximately achieved by solving the following bi-level optimization objective:
\begin{equation}
\label{CLP}
\centering
\begin{aligned}
    \delta &= \arg\max_{\delta\in \Delta} \mathcal{L}(\mathcal{I}^*, D_{tr}^{cla})\\
    &s.t. ~\mathcal{I}^* = \arg\min_{\mathcal{I}} \mathcal{L}(\mathcal{I}, PU_\delta(D_{tr}^{inv}))\\
\end{aligned}
\end{equation}
where $PU_\delta(D_{tr}^{inv})$ indicates that the output of sample $x\in D_{tr}^{inv}$ will be processed with perturbation $\delta$ according to Equation~\ref{inference}. In other words, $$\mathcal{L}(\mathcal{I}, PU_\delta(D_{tr}^{inv}))= \mathbb{E}_{x\in D_{tr}^{inv}} [\mathcal{R}(\mathcal{I}(PU_\delta(\mathcal{T}(x))), x)].$$  In addition, $\Delta$ denotes the search space which bounds the perturbations such that the perturbed confidence vectors $c'$ share the same maximal score position as $\mathcal{T}(x)$, hence preserving the classification accuracy of the target model.
\subsubsection{\textbf{Gradient Alignment}}
Unfortunately, directly solving the bi-level optimization problems in Equation~\ref{CLP} is intractable in deep learning scenarios because each iteration of the outer optimization requires the backpropagation computation in the inner learning problem. Thus, in the case of data poisoning attacks, a great number of works attempt to approximate the solution of the bi-level optimization~\cite{metapoison2020,witchbrew2021,munoz2017towards,yang2017generative}. We adapt the state-of-the-art attacks, Witches' Brew~\cite{witchbrew2021}, as our data poisoning strategy, which is effective. In particular, Witches' Brew~\cite{witchbrew2021} proposed to craft perturbations based on the guidance of gradient alignment to mislead the model training when the label of the training data cannot be changed. They aimed to craft poisoned data to mimic the gradient, which can achieve their adversarial objective (e.g., degrading the performance of the target model). These poisoned data can enforce the training gradient to align with the adversarial gradient. We adapt this method to the defense against model inversion attacks and we aim to align the training gradient w.r.t. the parameters of $\mathcal{I}$ on $PU(D_{tr}^{inv})$ (denoted as the poisoned gradient) with the defensive gradient on $D_{tr}^{cla}$ w.r.t. the parameters of $\mathcal{I}$ (denoted as the target gradient), which can be described formally:
\begin{equation}
\label{gradalign}
   \nabla\mathcal{L}_{def}(\mathcal{I}, D_{tr}^{cla}) \approx \nabla\mathcal{L}(\mathcal{I}, PU_\delta(D_{tr}^{inv})). 
\end{equation}

As a result, the minimization of $\mathcal{L}(\mathcal{I}, PU_\delta(D_{tr}^{cla}))$ on $D_{tr}^{inv}$ in the training of the inversion model will lead to the minimization of defensive loss $\mathcal{L}_{def}(\mathcal{I}, D_{tr}^{cla})$ on $D_{tr}^{cla}$, which improves the privacy of $D_{tr}^{cla}$ against model inversion attacks. Noting that, we need to fix the inversion model $\mathcal{I}$ to compute these two gradients, which requires pretraining a clean inversion model based on the attack data $D_{tr}^{inv}$ as the substitute before gradient computation. This is reasonable because the defender performing this gradient computation has access to attack data, that is produced by querying the target model.

To generate the perturbations satisfying Equation~\ref{gradalign}, we use the strategy from Witches' Brew~\cite{witchbrew2021} of matching the direction of the poisoned gradient and the target gradient instead of the gradient magnitudes, since aligning the gradient magnitudes might be infeasible. Thus, the negative cosine similarity is used to measure the direction alignment, and the optimized objective can be defined as follows:
\begin{equation}
\label{cos}
\small
\cos(\mathcal{I}, \delta)\triangleq 1 - \dfrac{\langle\nabla\mathcal{L}_{def}(\mathcal{I}, D_{tr}^{cla}),\nabla\mathcal{L}(\mathcal{I},PU_\delta(D_{tr}^{inv}))\rangle}{\left\|\nabla\mathcal{L}_{def}(\mathcal{I}, D_{tr}^{cla})\right\|\cdot\left\|\nabla\mathcal{L}(\mathcal{I}, PU_\delta(D_{tr}^{inv}))\right\|}.
\end{equation}

Based on a substitute inversion model $\mathcal{I}$, we can iteratively optimize $\cos(\mathcal{I}, \delta)$ to obtain a perturbation $\delta$ which can generate the inversion model with a small defensive loss. In order to enforce the constraints of confidence vectors, we normalized the perturbed confidence vectors using an arithmetic average strategy such that the sum of each confidence vector is $1$. The details of our LPA defense are outlined in Algorithm~\ref{alg:clp}. The first step is to compute the target gradient, which is the objective of gradient alignment (line 1). The perturbations are then initialized randomly (line 2). Post-process the confidence vectors $(\mathcal{T}(x)$ to the poisoned version $c'$ (line 4-5) and compute the alignment loss (line 6-7). Finally, update $\delta$ based on gradient descent strategy with the step size $\alpha_p$.

\begin{algorithm}
    \caption{LPA defense.}
    \label{alg:clp}
    \begin{algorithmic}[1]
        \REQUIRE A pre-trained substitute inversion model $\mathcal{I}$, Private data $D_{tr}^{cla}$, Attack data $D_{tr}^{inv}$, Initial perturbation $\delta_0$, and step size $\alpha_p$.
        \ENSURE Perturbations, $\delta$.
        
        \STATE Compute the target gradient $\nabla\mathcal{L}_{def}(\mathcal{I}, D_{tr}^{cla})$.
        \STATE $\delta = \delta_0$.
        
        \FOR{$i = 1$ to $R$ {\rm optimization rounds}}
            \STATE Post-process: $c' = \mathcal{T}(x) + \delta, \forall x\in D_{tr}^{inv}$.
            \STATE Normalize $c'$ using arithmetic average.
            \STATE Compute the poisoned gradient $\nabla\mathcal{L}(\mathcal{I}, PU_\delta(D_{tr}^{cla}))$.
            \STATE Compute the alignment loss $\cos(\mathcal{I}, \delta)$ in Eq.~\ref{cos}.
            \STATE Update $\delta$ using gradient descent with $\alpha_p$.
        \ENDFOR
        
        \RETURN $\delta$
    \end{algorithmic}
\end{algorithm}

It is important to note that the perturbations applied to the confidence vectors are not fixed, where each confidence vector is subjected to a unique perturbation. Specifically, for each confidence vector, a perturbation is initialized randomly. This random initialization ensures that the starting point for each perturbation is unique to the corresponding confidence vector. As a result, even before optimization, the perturbations are sample-specific. Although a batch of perturbations may be updated simultaneously, sharing an updating direction derived from the gradient descent, the initial randomization ensures that each perturbation remains tailored to the individual confidence vector.

\subsection{Method $2$: LFP defense}
Even though minor perturbations are added to the confidence vectors in the LPA defense, all confidence vectors are changed without any exception. As studied in classification tasks, only a small fraction of poisoned samples can also decrease the accuracy of the victim model by flipping the label of poisoned examples~\cite{yang2017generative,munoz2017towards,muller2020data}. Inspired by these works, we investigated the effectiveness of non-clean-label poisoning techniques against model inversion attacks in our scenario. Unlike LPA defense, the process of LFP can be divided into two phases. In the first phase, a picking strategy $S$ mentioned in Equation~\ref{inference} selects some sensitive confidence vectors as candidates for the poisoned examples. Then, in the second phase, we incorporate two attacks in regression~\cite{muller2020data} and classification tasks~\cite{witchbrew2021} to modify these selected candidates for the poisoned examples. Greater detail follows.

\subsubsection{\textbf{Selecting sensitive confidence vectors}} With limits on the number of examples that should be poisoned, it would be wise to select initial poisoned examples that will most effectively increase the impact of the data poisoning. It is expected that the weights of the inversion model will be sensitive enough to these poisoned confidence vectors that they can force these weights to produce low-quality reconstructions. Our strategy for selecting these effective candidates is based on their predicted labels and maximal scores. 

First, the number of classes involved in the poisoned dataset might influence their effectiveness in polluting the data. Intuitively, if all poisoned confidence vectors are selected from only one class of examples, it might be more difficult to impact the global performance of the victim model. Therefore, our strategy uniformly allocates candidates from different classes to maximize the poisoning effect. The second question is how we can identify the most effective confidence vectors from their counterparts in each class. One potential way is to categorize them based on their maximal scores. Most test samples will be assigned a confidence vector with a relatively high maximal score, implying the prediction is highly confident. These samples can be considered well-classified samples. Others will be considered poorly-classified samples that might lead to low confidence. Interestingly, we observed that the two sets of samples behave differently in a model inversion attack through experiments. 
It seems that the inversion model trained on well-classified confidence vectors delivered a lower reconstruction loss than the model trained on poorly-classified ones. Algorithm~\ref{alg:selecting} completely describes the procedure for selecting these sensitive confidence vectors.

\begin{algorithm}
    \caption{Select sensitive confidence vectors.}
    \label{alg:selecting}
    \begin{algorithmic}[1]
        \REQUIRE The clean dataset $D$, the budget (proportion of poisoned examples) $r$, and the pre-defined threshold of confidence $thr$.
        \ENSURE A poisoning candidature set $D_p$, and a set for remained clean examples: $D_c$.
        
        \STATE $D_p = [], D_c = []$
        \STATE Divide $D_{tr}^{inv}$ to multiple subsets according to the label: $D_{sub} = [D_1, D_2,..., D_l]$
        
        \FOR{$d$ in $D_{sub}$}
            \FOR{$c$ in $d$}
                \IF{{\rm max}($c$) $\geq thr$}
                    \STATE $D_{c}$.append($c$)
                \ELSE
                    \STATE $D_{p}$.append($c$)
                \ENDIF
            \ENDFOR
        \ENDFOR
        \IF{${\rm size}(D_p) < {\rm size}(D) \times r$}
            \STATE $k = {\rm size}(D) \times r - {\rm size}(D_p)$
            \STATE Transfer $k$ examples to $D_p$ from $D_c$
        \ENDIF
        \RETURN $(D_p, D_c)$
    \end{algorithmic}
\end{algorithm}

\subsubsection{\textbf{Produce poisoned examples}}
The extracted candidate dataset $D_p$ is then poisoned to produce poisoned confidence vectors, which will impact the reconstruction performance of the inversion models. The poisoning attacks used here cover different kinds of tasks. The flip poisoning attack in regression setting~\cite{muller2020data} is applied first, and, then the variant of Witches' Brew attack for classification models~\cite{witchbrew2021} is applied to further fine-tune the confidence vectors, which can improve the effectiveness of poisoned examples in the inversion tasks. The complete procedure of generating a poisoned example from one candidature in $D_p$ is described in Algorithm~\ref{alg:nclp}. 

Initially, we can select $D_p$ and $D_c$ with Algorithm~\ref{alg:selecting}. Based on the idea of a flip attack~\cite{muller2020data}, the rank of components in the initial poisoned sample is reversed (line $5$), denoted as $\hat{c}$. The second step (line $6-10$) is the same as the procedure of LPA. The produced perturbations $\delta$ are used to fine-tune $\hat{c}$ based on Witches' Brew method~\cite{witchbrew2021}. The modified $D_p$ and clean $D_c$ will construct the training data of the inversion model.

\begin{algorithm}
    \caption{LFP defense.}
    \label{alg:nclp}
    \begin{algorithmic}[1]
        \REQUIRE A pre-trained substitute inversion model $\mathcal{I}$, the private data $D_{tr}^{cla}$, attack data $D_{tr}^{inv}$, initial perturbation $\delta_0$, and step size $\alpha_p$.
        \ENSURE Perturbations, $\delta$
        \STATE Select $D_p$ and $D_c$ from $D_{tr}^{inv}$
        \STATE Compute the target gradient $\nabla\mathcal{L}_{def}(\mathcal{I}, D_{tr}^{cla})$
        \STATE $\delta = \delta_0$
        \FOR {$i = 1$ to $R$ optimization rounds}
            \STATE Reverse the $\hat{c} = Reverse(\mathcal{T}(x_p)), \forall x_p\in D_p$
            \STATE Post-process: $c' = \hat{c} + \delta, \forall x_p \in D_p$
            \STATE Normalize the $\delta$ using arithmetic average
            \STATE Compute the poisoned gradient $\nabla\mathcal{L}(\mathcal{I}, PU_\delta{D_p}+D_c)$
            \STATE Compute the alignment loss $\cos(\mathcal{I}, \delta)$ in Eq.~\ref{cos}
            \STATE Update $\delta$ with gradient descent
        \ENDFOR
        \RETURN $\delta$
    \end{algorithmic}
\end{algorithm}

\subsection{Discussion}
\subsubsection{Discussion of the LPA defense}
Like other defenses based on perturbations, our LPA also conceals privacy by perturbing the confidence vectors. The perturbations are effective against model inversion attacks, which has been validated in previous works~\cite{ye2022one}, but these perturbations might lead to poor utility~\cite{MID2020}. We believe that simple random perturbations contain excessive redundancy, and eliminating the redundant information contained in the perturbations will enhance the tradeoff between the utility of the target model and the defensive effectiveness. Our LPA defense employs a data poisoning method to generate the perturbations designed to increase reconstruction errors. In contrast to simple random perturbations, LPA perturbations primarily destroy the information relevant to the reconstruction tasks and will cause minor effects on other information. More specifically, previous perturbation-based methods generate random perturbations and are more likely to be untargeted. As a result, the produced perturbations are redundant and large, which makes the accuracy decrease significantly. In contrast, our method targets perturbing the privacy information that can be used by model inversion attacks and uses a gradient descent strategy to iteratively optimize the perturbation. The perturbations produced by our methods are smaller than previous perturbation-based defenses, which leads to a better trade-off and preserves the utility (i.e., classification accuracy) to the greatest degree possible. In addition, LPA only inserts perturbations to the confidence vectors. Therefore, LPA has not impacted the accuracy of the target model unless the perturbations are rather large. 

\subsubsection{LPA against adaptive adversaries} Our LPA remains resilient against adaptive attacks in which the attacker is aware of the LPA defense. Our defense strategy involves the uniform poisoning of confidence vectors, which means that all confidence vectors, regardless of whether an attack is present, are subjected to the poisoning process. This results in a scenario where the adversary receives $100\%$ poisoned confidence vectors during their queries, thereby eliminating the availability of any clean, normal data for comparative analysis.

\begin{figure}[!htbp]
    \centering
    \includegraphics[width=0.9\columnwidth]{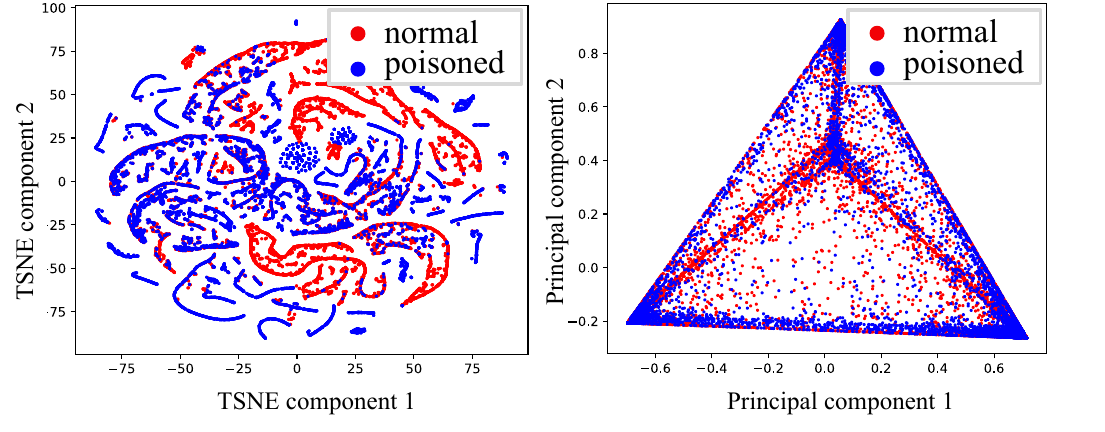}
    \caption{Visualization of the similarity between poisoned and normal confidence vectors based on t-SNE and PCA.}
    \label{fig:cluster}
\end{figure}

We also use unsupervised cluster methods, including t-SNE and PCA, to visualize the similarity between normal and poisoned confidence vectors. We have observed that unsupervised methods such as PCA and t-SNE do not provide a clear distinction between poisoned and normal confidence vectors (as shown in Figure~\ref{fig:cluster}). This further supports our claim that simple detection and discarding of poisoned data is not a viable strategy for circumventing our defense.

Furthermore, it is also impossible to recover the original confidence vector or clean gradients from the poisoned one. Even though adversaries are aware of the presence of LPA,  they only have access to the poisoned confidence vectors and do not receive any normal confidence vectors. As a result, they lack any information about the original confidence vector and clean gradients. Therefore, it is highly unlikely for an adversary to inverse the clean gradients. 
\subsubsection{Discussion of the LFP defense}
Unlike the LPA, accuracy will be affected by the LFP defense because the rank of initial poisoned confidence vectors is reversed, and therefore, the predicted labels are changed. Namely, $n\%$ poisoned data might lead to an approximate $n\%$ decrease in accuracy as a result of altering labels. To mitigate this degradation in accuracy, a technique can be adopted in this defense, i.e., swapping the positions of the maximal and the second-highest score before the reverse operation. And then keep the maximal score fixed and reverse other scores. Because poorly-classified data are preferred when selecting the initial poisoned examples, where samples leading to misclassification are more likely to be present, the second score might correspond to the correct label with a high probability. The swap operation replaces the original label with the second-most possible class, further improving the probability of correcting the misclassification before reversing. Therefore, the reverse operation aims to maximize their impact, while a swap operation is used to guarantee the accuracy of the target classifier. 

\subsubsection{Rationale of employing the pre-trained inversion model} The target gradient is expected to be calculated based on the attacker's model capable of reconstructing data with high quality. Nevertheless, it is generally intractable for defenders to engage directly in the learning process of an attacker's inversion model. In light of this, a pre-trained inversion model is intended to act as a proxy for the attacker's model, thereby simulating the attacker's behavior. To mimic the attacker's model, we ensure the pre-trained inversion model has learned the decoding rules of the confidence vectors and the representation of the reconstructed images. In the initial training stage of the attacker's model, we acknowledge that the gradient derived from our fixed inversion model may not immediately influence the attacker's model. However, as the attacker's model undergoes iterative optimization, the high-level features it captures begin to align with those of our fixed inversion model. This convergence suggests that the gradient, while initially less effective, becomes increasingly transferable. 

Generally speaking, our fixed model is able to capture the essential features that are critical to the reconstruction task. However, the gradients we introduce effectively prevent the attacker's model from converging to the state of our fixed model during the training process.

\section{Privacy Analysis}
Typically, a defense is successful if the reconstructed data of the attackers are not similar to the original data. To quantify the capability of our defenses based on data poisoning methods, we formally define the defense against model inversion attacks. 
\begin{Definition}($(\epsilon, \eta)$-Defense)
Let $PU$ be the post-processing unit of the defense, and $x'=\mathcal{I}(c)$ be the reconstructed version of the target data $x$ with confidence vector $c=\mathcal{T}(x)$, where $\mathcal{I}$ is trained on perturbed data pairs, $x$ and $PU(c)$, this defense is called a successful $(\epsilon, \eta)$-defense if the following property is held:
    \begin{equation}
    {\rm Pr}[\mathcal{R}(x', x)\geq\epsilon]\geq 1-\eta
    \end{equation}
\end{Definition}
Based on this definition, we quantify the capability of our LPA defense and investigate the relationship between the defensive effectiveness and the added noise in the following proposition. To simplify the presentation, the defensive loss $\mathcal{L}_{def}(\mathcal{I}(\mathcal{T}(x))), x)$ is denoted as $\mathcal{L}_{def}(\mathcal{I})$, which is computed on $D_{tr}^{cla}$. Similarly, we denote $\mathcal{L}(\mathcal{I}(PU_\delta(\mathcal{T}(x))), x)$ as $\mathcal{L}(\mathcal{I},\delta)$, which is computed on $D_{tr}^{inv}$. And the gradient similarity between $\nabla\mathcal{L}_{def}(\mathcal{I}^k)$ and $\nabla\mathcal{L}(\mathcal{I}^k)$ is denoted as $\cos(\mathcal{I}^k, \delta)\triangleq \dfrac{\langle\nabla\mathcal{L}_{def}(\mathcal{I}^k),\nabla\mathcal{L}(\mathcal{I}^k,\delta)\rangle}{\left\|\nabla\mathcal{L}_{def}(\mathcal{I}^k, \delta)\right\|\cdot\left\|\nabla\mathcal{L}(\mathcal{I}^k,\delta)\right\|}$, where $\mathcal{I}^k$ indicates the inversion model at iteration $k$. 
\begin{Proposition}
Let $\mathcal{L}_{def}(\mathcal{I})$ have a Lipschitz continuous gradient with constant $L>0$, and $\alpha$ be the learning rate of training inversion model. We assume that the reconstruction loss on the target data of a randomly initialized inversion model (i.e., without training) follows normal distribution $\mathcal{N}(\epsilon_0, \sigma^2)$. If $$\alpha<\cos(\mathcal{I}^k,\delta)\dfrac{l\left\|\mathcal{L}_{def}(\mathcal{I}^k)\right\|}{L\left\|\mathcal{L}(\mathcal{I}^k, \delta)\right\|}$$ for constant $c<1$ and any iteration $k$, then $$\mathcal{L}_{def}(\mathcal{I}^{k+1})<\mathcal{L}_{def}(\mathcal{I}^0)-\epsilon_1(k,\delta)$$ where $$\epsilon_1(k, \delta)=\dfrac{l}{L}\sum_{j=0}^k \left\|\nabla\mathcal{L}_{def}(\mathcal{I}^j)\right\|^2\cos^2(\mathcal{I}^j,\delta).$$
Further, if we define the reconstruction error of the inversion model trained $k$ iterations under defense $PU_\delta$ as $\mathcal{R}_k(x', x) = \mathcal{L}(\mathcal{I}^k, \delta) = \mathcal{L}(\mathcal{I}^k(PU_\delta(\mathcal{T}(x))),x)$, it satisfies:
$${\rm Pr}[\mathcal{R}_{k+1}(x', x)>\epsilon_0]\geq 1- \Phi(\dfrac{-\epsilon_1(k, \delta)}{\sigma})$$where $x'=\mathcal{I}(PU_\delta(\mathcal{T}(x)))$ is the reconstructed data and $\Phi(t)=\dfrac{1}{\sqrt{2\pi}}\int_{-\infty}^t exp(\dfrac{-z^2}{2})dz$ is the cumulative distribution function of standard normal distribution.
\end{Proposition}
\textbf{\emph{Proof of Proposition} 1.} Consider the inversion model trained perturbed confidence vectors with gradient descent update $\mathcal{I}^{k+1}=\mathcal{I}^{k}-\alpha\nabla\mathcal{L}(\mathcal{I}^k,\delta)$. We can estimate the defensive loss $\mathcal{L}_{def}(\mathcal{I}^{k+1})$ at $k+1$ iteration based on the Lipschitz smoothness:
{\small
\begin{equation}
\begin{aligned}
    \mathcal{L}_{def}(\mathcal{I}^{k+1})\leq&\mathcal{L}_{def}(\mathcal{I}^k)-\langle\alpha\nabla\mathcal{L}_{def}(\mathcal{I}^k),\nabla\mathcal{L}(\mathcal{I}^k,\delta)\rangle\\&+\alpha^2L\left\|\nabla\mathcal{L}(\mathcal{I}^k,\delta)\right\|
\end{aligned}
\end{equation}}
Expand the inner product of two gradients:
{\small $$\langle\nabla\mathcal{L}_{def}(\mathcal{I}^k),\nabla\mathcal{L}(\mathcal{I}^k,\delta)\rangle=\left\|\nabla\mathcal{L}_{def}(\mathcal{I}^k)\right\|\cdot\left\|\nabla\mathcal{L}(\mathcal{I}^k,\delta)\right\|\cos(\mathcal{I}^k,\delta)$$}
As a result,{\small 
\begin{equation}
\begin{aligned}
    \mathcal{L}_{def}(\mathcal{I}^{k+1})\leq&\mathcal{L}_{def}(\mathcal{I}^k)\\&-\alpha\left\|\nabla\mathcal{L}_{def}(\mathcal{I}^k)\right\|\cdot\left\|\nabla\mathcal{L}(\mathcal{I}^k,\delta)\right\|\cos(\mathcal{I}^k,\delta)\\&+\alpha^2L\left\|\nabla\mathcal{L}(\mathcal{I}^k,\delta)\right\|\\
    \leq&\mathcal{L}_{def}(\mathcal{I}^k)\\&-\alpha(\dfrac{\left\|\nabla\mathcal{L}_{def}(\mathcal{I}^k)\right\|}{\left\|\nabla\mathcal{L}(\mathcal{I}^k,\delta)\right\|}\cos(\mathcal{I}^k,\delta)-\alpha L)\left\|\nabla\mathcal{L}(\mathcal{I}^k,\delta)\right\|^2\\
    <&\mathcal{L}_{def}(\mathcal{I}^k) - \alpha^2 L\left\|\nabla\mathcal{L}(\mathcal{I}^k,\delta)\right\|^2
\end{aligned}
\end{equation}}
Based on the assumption on the learning rate $\alpha<\cos(\mathcal{I}^k,\delta)\dfrac{l\left\|\mathcal{L}_{def}(\mathcal{I}^k)\right\|}{L\left\|\mathcal{L}(\mathcal{I}^k, \delta)\right\|}$, we can reveal that:
{\small
\begin{equation}
\begin{aligned}
    \mathcal{L}_{def}(\mathcal{I}^{k+1})<&\mathcal{L}_{def}(\mathcal{I}^k)-\left\|\nabla\mathcal{L}_{def}(\mathcal{I}^k)\right\|^2\dfrac{l\cdot\cos^2(\mathcal{I}^k, \delta)}{L}
\end{aligned}
\end{equation}}
Compute the sum over all iterations:
{\small
\begin{equation}
    \begin{aligned}
        \mathcal{L}_{def}(\mathcal{I}^{k+1})-\mathcal{L}_{def}(\mathcal{I}^{0})<-\dfrac{l}{L}\sum_{j=0}^k{\left\|\nabla\mathcal{L}_{def}(\mathcal{I}^j)\right\|^2}\cos^2(\mathcal{I}^j, \delta)
    \end{aligned}
\end{equation}}
Recall the definition of defensive loss and the reconstruction error at $k+1$-th iteration:
{\small
\begin{equation}
    \begin{aligned}
        \mathcal{L}_{def}(\mathcal{I}^{k+1}(PU_\delta(c)), x) &= - \mathcal{L}(\mathcal{I}^{k+1}(PU_\delta(c)),x)\\
        &= - \mathcal{R}_{k+1}(PU_\delta(c),x)\\
        &= - \mathcal{R}_{k+1}(x',x)
    \end{aligned}
\end{equation}}
Therefore,
{\small
\begin{equation}
    \begin{aligned}
        \mathcal{R}_{k+1}(x',x)-\mathcal{R}_0(x',x)>\dfrac{l}{L}\sum_{j=0}^k{\left\|\nabla\mathcal{L}_{def}(\mathcal{I}^j)\right\|^2}\cos^2(\mathcal{I}^j, \delta)
    \end{aligned}
\end{equation}}
As $\mathcal{I}^0$ is a randomly initialized model without training, $\mathcal{R}_0(x',x) = \mathcal{L}(\mathcal{I}^0)$ follows $\mathcal{N}(\epsilon_0, \sigma^2)$. We may consider the bound of $\mathcal{L}(\mathcal{I}^{k+1})$ based on $\epsilon_0$. If we define {\small$$\epsilon_1(k, \delta)=\dfrac{l}{L}\sum_{j=0}^k{\left\|\nabla\mathcal{L}_{def}(\mathcal{I}^j)\right\|^2}\cos^2(\mathcal{I}^j, \delta),$$} the upper bound of the reconstruction error at $k+1$-th iteration $\mathcal{R}_{k+1}(x',x)$ is $\epsilon \sim \mathcal{N}(\epsilon_0+\epsilon_1(k, \delta), \sigma^2)$. And then, the probability of this reconstruction error being greater than $\epsilon_0$ can be computed as follows:
{\small$${\rm Pr}[\mathcal{R}_{k+1}(x',x)>\epsilon_0]\geq1-\Phi(\dfrac{-\epsilon_1(k,\delta)}{\sigma}).$$}

\section{Experiments}
\label{sec:exp}
\subsection{Experimental setup}
\subsubsection{Model Training}
In our experiments, the full training data of these datasets were divided into four parts: the private data for the classifier, i.e., $D_{tr}^{cla}$ and $D_{test}^{cla}$, and the attack data for inversion model, i.e., $D_{tr}^{inv}$ and $D_{test}^{inv}$. In particular, $D_{tr}^{cla}$ is used to train the target classifier model. And the attacker uses $D_{tr}^{inv}$ to train their inversion model. With MNIST, we assumed that the attacker had accessed $5$ classes of images (i.e., digit $5-9$) to conduct inversion tasks, while another $5$ classes of private images (digit $0-4$) are the training data of the target classifier. Likewise, Fashion-MNIST is also divided into private data with $5$ classes and attack data with another $5$ classes, with no overlap between private and attack data. With the FaceScrub dataset, we selected photos of $100$ people from FaceScrub as the private data, which we used to train the target classifier, and photos of another $100$ people as the public data, which we used to train the inversion model.

\subsubsection{Model Architecture}
Our focus in the experiments was the knowledge alignment attacks introduced in~\cite{DBLP:journals/corr/abs-1902-08552}. Therefore, we followed the corresponding architectures of an encoder/decoder comprising a victim CNN classifier model as the encoder and the inversion model as the decoder. The classifiers consisted of CNN blocks, two fully-connected layers, and a softmax layer used to transform the extracted features into a valid confidence vector. Specifically, the classifiers for MNIST and Fashion-MNIST used three CNN blocks to extract the features, while four CNN blocks were used in the classifier for FaceScrub due to the complexity of face images.

The inversion model can be considered as the inverse of the classifier, but it only consisted of transposed CNN blocks. For MNIST and Fashion-MNIST, the inversion model consisted of four transposed CNN blocks, while the FaceScrub inversion model used five transposed CNN blocks. 

\subsubsection{Evaluation Metrics}
We used two quantitative measurements to evaluate the utility loss: accuracy and confidence score distortion. We argue that accuracy is not a fully precise indicator of utility. We have also used the confidence score distortion caused by defense mechanisms to quantitatively assess the utility loss. The distortion is roughly defined as the Euclidean distance between the confidence vectors from the original classifier without a defense and the defended model. Moreover, to evaluate the defense strength and privacy leakage, we measured the mean squared error between the original data and its reconstructed result (i.e., reconstruction error) as well as the classification accuracy of the reconstructed data (referred to as attack accuracy). 
Reconstruction error measures the pixel-wise similarity between the original and reconstructed images. While it provides a quantitative assessment, it does not always align perfectly with human perception of similarity. Attack accuracy reflects the similarity between the reconstructed and original images from the perspective of a machine learning model, specifically, whether the classifier can still categorize the reconstructed image into the same class as the original. Both metrics have their limitations when used in isolation. The reconstruction error offers a pixel-level assessment, whereas attack accuracy provides a measure of the semantic information contained in the image, rather than just the pixel information. It complements the reconstruction error by considering the class-discriminative features recognized by the classifier. 

\subsection{Experimental results}
\subsubsection{The effectiveness of data poisoning} 
\begin{figure*}[!t]
    \centering
    \includegraphics[width=0.8\linewidth]{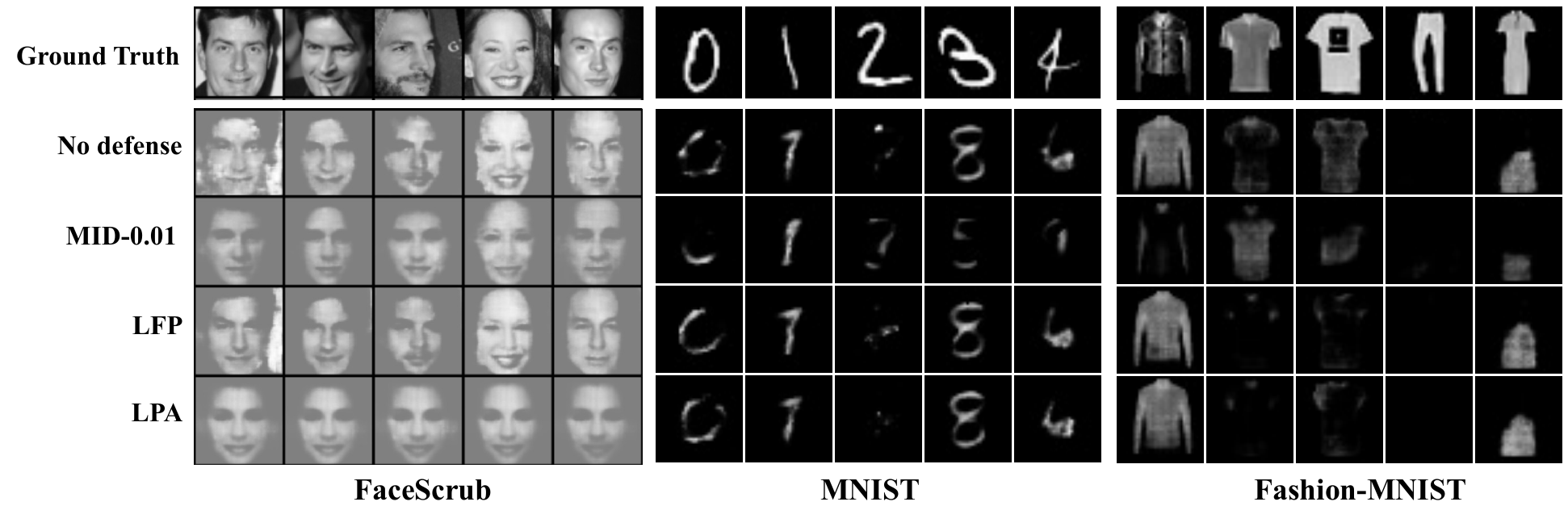}
    \caption{The reconstructed images generated by inversion models under different defense strategies.}
    \label{fig:diff_methods}
\end{figure*}

Our first experiment was designed to examine the effectiveness of data poisoning as a defense against a model inversion attack. The reconstructed images produced by the different inversion models are shown in Fig.~\ref{fig:diff_methods}. The first row shows the ground truth images, and the second row shows the reconstructed images from a separate inversion model~\cite{DBLP:journals/corr/abs-1902-08552} with no defenses. The third row depicts the reconstructed pictures generated by an inversion model trained with a MID-defended classifier, with the parameter $\lambda$ determining the trade-off between mutual information and classification error set to $0.01$. The following two rows contain the reconstruction results of the inversion model compromised by our defensive techniques, LPA and LFP. On the FaceScrub dataset, we can see that an inversion model without defenses can effectively recover the original image, whereas the LFP technique had no effect on the quality of reconstructed images and failed to protect the privacy of the original images. However, our LPA drastically degraded reconstruction quality. Under our LPA defense, it is impossible to determine an individual's identity from the reconstructed images. Compared to the baseline, MID~\cite{MID2020}, our LPA method provides superior privacy protection for all samples, while MID may fail for particular data that are simple to reconstruct. 

Table~\ref{tab:accuracy_diff} also displays the quantifiable outcomes of privacy leaks, including the reconstruction errors and the attack accuracy. 
Consistent with the visual results, the improvement in reconstruction error generated by the LFP strategy is not readily apparent. However, our LPA defense can greatly increase the reconstruction error and reduce the attack accuracy of model inversion attacks despite requiring no retraining. We can conclude that our LPA outperforms the SOTA MID based on retraining on FaceScrub. While mitigating the threat of model inversion, all defensive measures inevitably impact the utility of the target classifier. 

Based on two metrics related to the utility of the classifier (i.e., model accuracy and the confidence score distortion), we design an extensive set of experiments to explore the side effects of the defensive methods. Accuracy describes the ability of the target model to make predictions in classification tasks. The difference in confidence vectors can be considered as a loss of utility of the target classifier. As shown in Table~\ref{tab:accuracy_diff}, LFP is impractical because the changing label strategy greatly degrades the classification accuracy of the target model despite the minor confidence score distortion. By contrast, our LPA method causes zero loss of the classification accuracy of the target model and less confidence score distortion than the MID and LFP methods. 

The outcomes on MNIST, FashionMNIST, and CIFAR-10 datasets mirror those observed with FaceScrub. \textcolor{black}{The shift in data distribution for MNIST and FashionMNIST is more pronounced, which, in turn, restricts the {\em visual quality} of reconstructed images. This explains why the reconstructed images on FaceScrub seem better. However, it would not affect the effectiveness of two quantitative metrics (Reconstruction error and Attack accuracy) in evaluating the model inversion attacks as illustrated in the experimental results.} Our experiments reveal that our LPA defense strategy remains more robust and offers greater utility compared to the MID approach.  

Taking our investigations a step further, we analyzed the reasons why LFP defense is ineffective. First, as shown in back-gradient attacks~\cite{munoz2017towards}, deep neural networks are more resilient against poisoning attacks that manipulate only a very small fraction of training data when compared to other learning algorithms. Even if our LFP modifies the labels of the poisoned samples, the inversion model is still able to learn how to invert the data from an abundance of clean data. This presents the challenge of poisoning a tiny subset of training data to undermine the functionality of the inversion model. Moreover, it may be challenging to assess the efficacy of each poisoning candidate. Consequently, we are not able to pick the most effective subset from the complete training data. This issue limits the performance of defenses in comparison to our LPA, which corrupts all data. 
In the rest of the experiments, we will focus on the evaluation of our LPA defense while skipping the LFP due to its ineffectiveness of LFP.

\begin{table}[!t]
\caption{Utility and privacy of the target model under different defenses. The utility metrics include model accuracy (Model Acc.) and confidence score distortion (Conf. Dist.), and privacy metrics include attack accuracy (Attack Acc.) and reconstruction error (Recon. Error). LFP and LPA are our proposed defenses based on data poisoning with different poisoning strategies. \label{tab:accuracy_diff}}
\centering
\renewcommand{\arraystretch}{1.5}
\begin{tabular}{|c|c|cc|cc|}
\hline
\multirow{2}{*}{\textbf{Dataset}}&\multirow{2}{*}{\textbf{Methods}}&\multicolumn{2}{c|}{\textbf{Utility}}&\multicolumn{2}{c|}{\textbf{Privacy}}\\
\cline{3-6}
&&\makecell[c]{Model\\Acc.}&\makecell[c]{Conf.\\Dist.}&\makecell[c]{Attack\\Acc.}&\makecell[c]{Recon.\\Error}\\
\hline
\hline
\multirow{3}{*}{MNIST}&No-defense & $99.98\%$&-&$94.80\%$&$0.88357$\\
&MID-0.01 &$99.86\%$&$0.01492$&$83.99\%$&$0.88405$\\
&LFP &$94.96\%$&$0.06998$&$96.53\%$&$0.88295$\\
&LPA &$\textbf{99.98\%}$&$\textbf{0.00014}$&$\textbf{74.72\%}$&$\textbf{0.88915}$\\
\hline
\multirow{3}{*}{\makecell[c]{Fashion-\\MNIST}}&No-defense & $94.72\%$&-&$65.76\%$&$0.60050$\\
&MID-0.01 &$94.26\%$&$0.06957$&$40.44\%$&$0.61173$\\
&LFP &$92.11\%$&$0.05289$&$46.62\%$&$0.60179$\\
&LPA &$\textbf{94.72\%}$&$\textbf{0.00021}$&$\textbf{40.07\%}$&$\textbf{0.62192}$\\
\hline
\multirow{3}{*}{FaceScrub}&No-defense & $94.82\%$&-&$32.32\%$ &$0.2076$\\
&MID-$0.01$ &$93.94\%$&$0.11463$&$25.64\%$ &$0.2274$\\
&LFP & $90.57\%$&$0.06038$&$30.28\%$&$0.2140$\\
&LPA & $\textbf{94.82\%}$&$\textbf{0.00340}$&$\textbf{5.08\%}$ &$\textbf{0.2318}$\\
\hline
\multirow{3}{*}{CIFAR10}&No-defense & $86.20\%$&-&$46.10\%$ &$0.8662$\\
&MID-$0.001$ &$85.96\%$&$0.20209$&$40.48\%$ &$0.8828$\\
&LFP & $84.80\%$&$0.04913$&$43.90\%$&$0.8625$\\
&LPA & $\textbf{86.20\%}$&$\textbf{0.00021}$&$\textbf{38.64\%}$ &$\textbf{0.8898}$\\
\hline
\end{tabular}
\end{table}

\subsubsection{Impact of the loss function} \textcolor{black}{Given the black-box nature of interactions between the defender and the inversion model, the attacker's choice of loss function remains unknown during defense strategy development. This motivated us to investigate how different loss function selections affect the defense mechanism's effectiveness. In the context of model inversion attacks, two metrics are commonly used to guide image reconstruction: Mean Squared Error (MSE) and Peak Signal-to-Noise Ratio (PSNR). Our experiments examine various scenarios where both attacker and defender independently select one of these loss functions, with the defender having no knowledge of the attacker's choice. Table~\ref{tab:loss_func_face} presents detailed results from these experiments on the FaceScrub dataset.}
\begin{table}[!t]
\caption{Utility and privacy of the target model under diverse loss functions on FaceScrub. `Defe. loss' refers to the loss function for the pre-trained inversion model. `Atta. loss' denotes the loss function of attacker's inversion model.
\label{tab:loss_func_face}}
\centering
\renewcommand{\arraystretch}{1.5}
\begin{tabular}{|c|c|cc|cc|}
\hline
\multirow{2}{*}{\textbf{Atta.}}&\multirow{2}{*}{\textbf{Defe.}}&\multicolumn{2}{c|}{\textbf{Utility}}&\multicolumn{2}{c|}{\textbf{Privacy}}\\
\cline{3-6}
\textbf{loss}&\textbf{loss}&\makecell[c]{Model\\Acc.}&\makecell[c]{Conf.\\Dist.}&\makecell[c]{Attack\\Acc.}&\makecell[c]{Recon.\\Error}\\
\hline
\hline
\multirow{3}{*}{MSE}&No-defense & $94.82\%$&-&$32.32\%$&$0.2076$\\
&MSE &$94.82\%$&$\textbf{0.00340}$&$5.08\%$&$0.2318$\\
&PSNR &$94.82\%$&$0.00363$&$5.04\%$&$0.2335$\\
\hline
\multirow{3}{*}{\makecell[c]{PSNR}}&No-defense & $94.82\%$&-&$31.17\%$&$0.2104$\\
&MSE &$94.82\%$&$\textbf{0.00340}$&$\textbf{4.06\%}$&$\textbf{0.2321}$\\
&PSNR &$94.82\%$&$0.00363$&$4.63\%$&$0.2316$\\
\hline
\end{tabular}
\end{table}

\textcolor{black}{Table~\ref{tab:loss_func_face} demonstrates that our approach maintains its effectiveness across different loss functions used by the attacker, regardless of whether the attacker employs MSE or PSNR as the inversion model's loss function. This enhancement is evidenced by consistently lower attack accuracy and higher reconstruction error. The defensive performance remains robust even when the defender's loss function differs from the attacker's choice.}
\textcolor{black}{The results also reveal that implementing effective defenses becomes more straightforward when attackers use PSNR to train their attack models. This observation can be attributed to two factors. First, MSE typically yields better performance than PSNR in training attack models. Second, since MSE serves as the evaluation metric for image reconstruction quality in our experiments, attackers using MSE naturally achieve lower reconstruction errors.}
\subsubsection{Impact of attack data} \textcolor{black}{The defender lacks knowledge about which specific query samples are used to train the attack model, despite these samples being from query data. We investigated how this uncertainty affects our defensive strategies by partitioning the inversion model's training data ($D_{tr}^{inv}$) into two distinct subsets: $D_1^{inv}$ and $D_2^{inv}$. Our experiment simulates independent subset selection by both parties, where the defender determines the training data for the pretrained inversion model, and the attacker's selection defines the training set for the attack model. The experimental outcomes on FaceScrub are presented in Table~\ref{tab:atta_data_face}.}

\begin{table}[!t]
\caption{Utility and privacy of the target model under different attack data on FaceScrub. `Defe. data' refers to the training data for the pre-trained inversion model. `Atta. data' denotes the training data of the inversion model of the attacker. 
\label{tab:atta_data_face}}
\centering
\renewcommand{\arraystretch}{1.5} 
\begin{tabular}{|c|c|cc|cc|}
\hline
\multirow{2}{*}{\textbf{Atta.}}&\multirow{2}{*}{\textbf{Defe.}}&\multicolumn{2}{c|}{\textbf{Utility}}&\multicolumn{2}{c|}{\textbf{Privacy}}\\
\cline{3-6}
\textbf{data}&\textbf{data}&\makecell[c]{Model\\Acc.}&\makecell[c]{Conf.\\Dist.}&\makecell[c]{Attack\\Acc.}&\makecell[c]{Recon.\\Error}\\
\hline
\hline
\multirow{3}{*}{$D_1^{inv}$}&No-defense & $94.82\%$&-&$30.01\%$&$0.2115$\\
\cline{2-6}
&$D_1^{inv}$ &$94.82\%$&$\textbf{0.00302}$&$4.80\%$&$\textbf{0.2256}$\\
\cline{2-6}
&$D_2^{inv}$ &$94.82\%$&$0.00316$&$5.16\%$&$0.2242$\\
\hline
\multirow{3}{*}{\makecell[c]{$D_2^{inv}$}}&No-defense & $94.82\%$&-&$32.32\%$&$0.2103$\\
\cline{2-6}
&$D_1^{inv}$ &$94.82\%$&$\textbf{0.00302}$&$4.16\%$&$0.2149$\\
\cline{2-6}
&$D_2^{inv}$ &$94.82\%$&$0.00316$&$\textbf{3.79\%}$&$0.2153$\\
\hline
\end{tabular}
\end{table}

\textcolor{black}{In this experiment, $D_1^{inv}$ and $D_2^{inv}$ are mutually exclusive subsets, each containing $14702$ images. Our approach significantly enhances privacy protection without compromising the utility of the defended model with different subset. Moreover, even when the defender's pre-training data differs from the attacker's choice of attack data, our defense mechanism continues to provide strong privacy protection. The observed reconstruction error and attack accuracy are comparable to those observed when both parties operate on the same dataset. For instance, when the defender chooses $D_2^{inv}$ to pre-train the substitute inversion model, the attacker might employ a different dataset (i.e., $D_1^{inv}$) to train the inversion model. In this case, the reconstruction error can escalate to $0.2242$, which is greatly higher than the error observed in the absence of defenses (i.e., $0.2115$).}

\subsubsection{Effects of model architectures}
\begin{figure}[!t]
    \centering
    \includegraphics[width=0.7\linewidth]{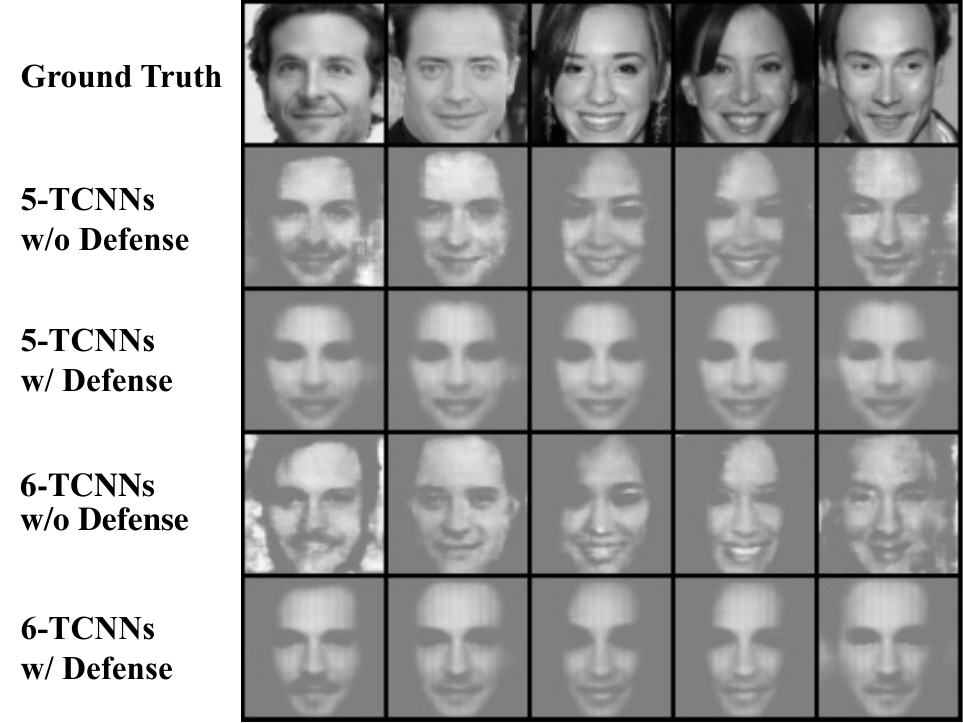}
    \caption{The reconstructed images generated by inversion models with different architectures. `w/' indicates `with LPA defense' while `w/o' signifies `without LPA defense'.}
    \label{fig:transfer}
\end{figure}

Our defense strategy generates poisoned samples using a substitute model as a proxy for the attacker's inversion model. The effectiveness of this LPA defense mechanism depends on the transferability of poisoned samples across different model architectures. To evaluate this aspect, we conducted experiments using a substitute model comprising 5 transposed CNN blocks ($5$-TCNNs) to generate perturbations, which serve as poisoned samples to contaminate the attacker's training data. We trained two inversion models with $5$ and $6$ transposed CNN blocks based on the polluted data. As a benchmark for effectiveness, we also trained two inversion models with no defenses. Fig.~\ref{fig:transfer} demonstrates the cross-architecture transferability of our poisoned samples from $5$-TCNNs to $6$-TCNNs, despite a significant decrease in efficacy. Specifically, under our LPA defense, the $5$-TCNN inversion model produces completely unrecognizable reconstructions, while the $6$-TCNN model's reconstructions only reveal gender information. Importantly, the primary private information, i.e., identity, remains safeguarded.

In addition, Table~\ref{tab:transfer} quantitatively demonstrates the degraded efficacy of our LPA defense across different architectures. The inversion model with $6$-TCNNs yields a higher attack accuracy than its counterpart with $5$-TCNNs, but it is significantly less than the attack accuracy with no defense. Although the attack accuracy of $20.46\%$ against $6$-TCNNs might not be acceptable for the identity classification task of $100$ identities, our defense mechanism prioritizes balancing privacy protection with model utility. 
It significantly reduces the success rate of the attacks while maintaining the utility.
\begin{table}[!t]
\caption{The performance of our LPA defense when transferred to inversion models with different architectures on FaceScrub dataset. `w/' indicates `with LPA defense' while `w/o' signifies `with out LPA defense'. \label{tab:transfer}}
\centering
\renewcommand{\arraystretch}{1.3} 
\begin{tabular}{|c|c|c|cc|}
\hline
\multirow{2}{*}{\textbf{Substitute}}&\multirow{2}{*}{\textbf{Architecture}}&\multirow{2}{*}{\textbf{Defense}}&\multicolumn{2}{c|}{\textbf{Privacy}}\\
\cline{4-5}
&&&\makecell[c]{Attack\\Acc.}&\makecell[c]{Recon.\\Error}\\
\hline
\hline
\multirow{4}{*}{{\scriptsize $5$-TCNNs}}&$5$-TCNNs &w/o&$32.32\%$ &$0.2076$\\
&$5$-TCNNs &w/&$\textbf{5.08\%}$ &$\textbf{0.2312}$\\
&$6$-TCNNs &w/o&$33.67\%$ &$0.2078$\\
&$6$-TCNNs &w/& $20.46\%$ &$0.2306$\\
\hline
\multirow{2}{*}{{\scriptsize $6$-TCNNs}}
&$5$-TCNNs &w/&$21.95\%$ &$0.2269$\\
&$6$-TCNNs &w/& $\textbf{13.11\%}$ &$\textbf{0.2298}$\\
\hline
\end{tabular}
\end{table}

\begin{figure}[!h]
    \centering
    \includegraphics[width=0.9\linewidth]{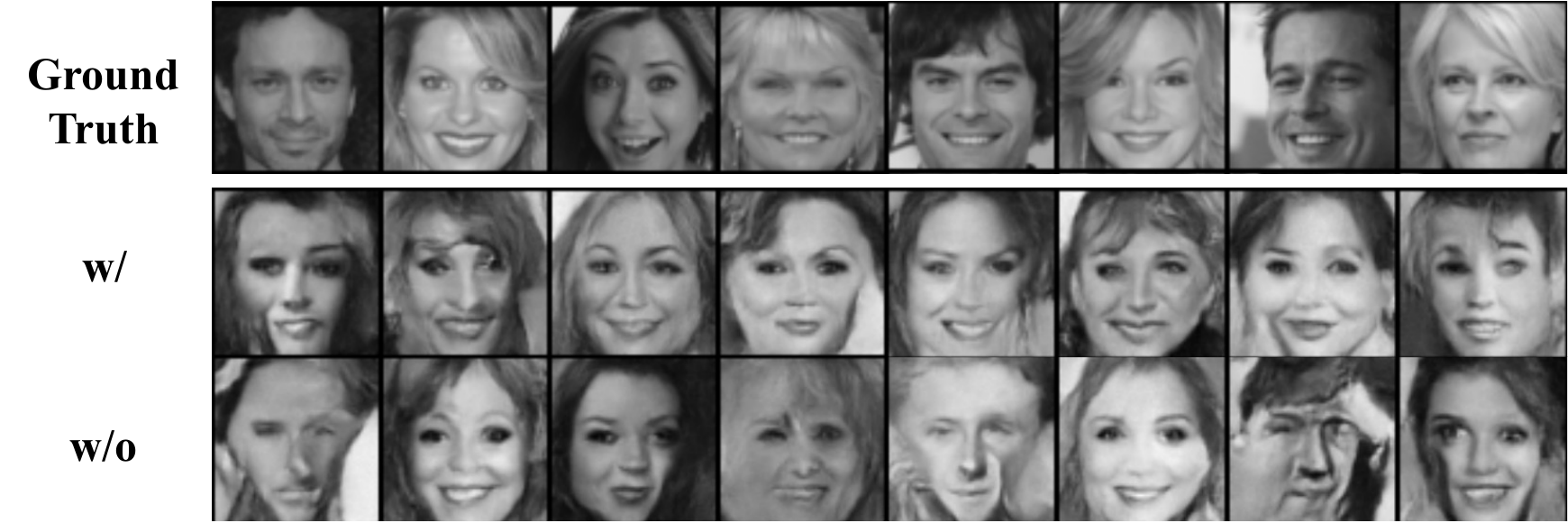}
    \caption{The reconstructed images on FaceScrub generated by inversion models with diffusion model architectures. The generative guidance is tuned to $0.5$.}
    \label{fig:diffusion_face}
\end{figure}

\textcolor{black}{We also explored the potential of diffusion models in reconstructing data from confidence vectors by developing a conditional diffusion model that generates data from random noise, conditioned by a confidence vector. This innovative framework exhibits an enhanced capacity to reconstruct data from confidence vectors, potentially posing amplified privacy risks. Figure~\ref{fig:diffusion_face} demonstrates the natural appearance of the generated images, achieved with a generative guidance strength of $0.5$. Although our defensive measures maintain reconstruction quality, the undefended scenario produces reconstructions that more accurately reflect the original images' characteristics, notably preserving the gender attribute. Moreover, quantitative analysis (Table~\ref{tab:diffusion}) reveals that while our defensive measures successfully reduce attack accuracy, they unexpectedly result in lower reconstruction error. This limited effectiveness can be attributed to the substantial architectural differences between the substitute TCNN inversion model and the diffusion-based attack model, which reduces the transferability and impact of our poisoned examples.}

In general, our LPA defense is still effective in mitigating privacy risks despite the attacker's inversion model having a different number of TCNN blocks than the substitute model. However, if the attacker selects an architecture completely different from that of the substitute model, the efficacy of our LPA could be compromised. In the future, we intend to incorporate the ensemble strategy to improve the performance and robustness of our LPA defense, ensuring it remains robust against inversion models with varying architectures.

\begin{table}[!t]
\caption{The performance of our LPA defense when transferred to inversion models with conditional diffusion model architectures on MNIST dataset. `Guidance' indicates the strength of generative guidance. \label{tab:diffusion}}
\centering
\renewcommand{\arraystretch}{1.3} 
\begin{tabular}{|c|c|c|cc|}
\hline
\multirow{2}{*}{\textbf{Architecture}}&\multirow{2}{*}{\textbf{Guidance}}&\multirow{2}{*}{\textbf{Defense}}&\multicolumn{2}{c|}{\textbf{Privacy}}\\
\cline{4-5}
&&&\makecell[c]{Attack\\Acc.}&\makecell[c]{Recon.\\Error}\\
\hline
\hline
&$0$&w/o&$73.20\%$ &$0.4275$\\
\textbf{Conditional}&$0$ &w/&$\textbf{66.01\%}$ &$0.4057$\\
\textbf{Diffusion}&$0.5$&w/o&$79.32\%$ &$0.4275$\\
\textbf{Model}&$0.5$&w/& $68.62\%$ &$0.4029$\\
&$2.0$&w/o&$83.21\%$ &$\textbf{0.4275}$\\
&$2.0$&w/& $72.84\%$ &$0.4145$\\
\hline
\end{tabular}
\end{table}

\subsubsection{Effects of the magnitude of perturbations} 
\begin{figure}[!t]
    \centering
    \includegraphics[width=0.6\linewidth]{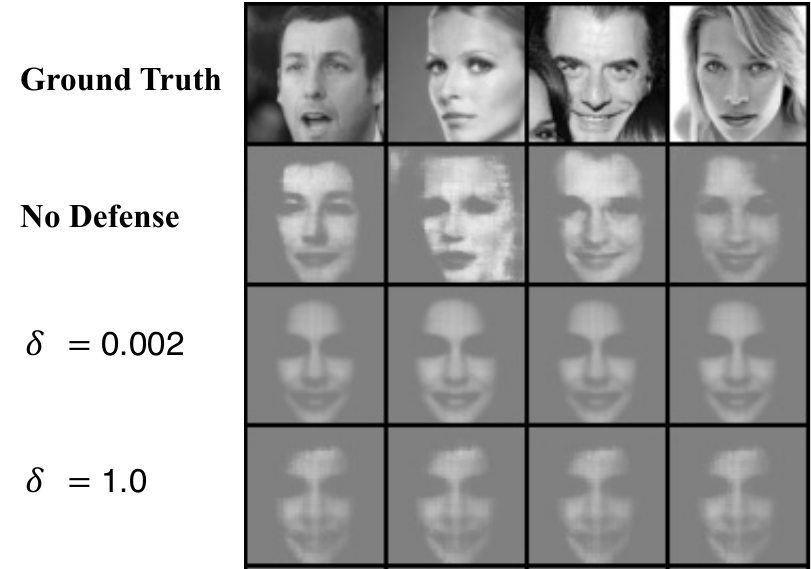}
    \caption{The reconstructed images generated by inversion models using the different magnitudes of perturbations.}
    \label{fig:magnitude}
\end{figure}

Our subsequent experiments aimed to investigate the effects of perturbation magnitude on the effectiveness of our LPA defense. We used the $\ell_2$ norm to quantify the magnitude of perturbations integrated into the confidence vectors.
\begin{table}[!t]
\caption{The performance of our LPA defense with different perturbation magnitudes on FaceScrub. \label{tab:magnitude}}
\centering
\begin{tabular}{|c|cc|cc|}
\hline
\multirow{2}{*}{\textbf{$\ell_2(\delta)$}}&\multicolumn{2}{c|}{\textbf{Utility}}&\multicolumn{2}{c|}{\textbf{Privacy}}\\
\cline{2-5}
&\makecell[c]{Model\\Acc.}&\makecell[c]{Conf.\\Dist.}&\makecell[c]{Attack\\Acc.}&\makecell[c]{Recon.\\Error}\\
\hline
\hline
$0$ & $94.82\%$&-&$32.32\%$ &$0.2076$\\
$0.002$ &$\textbf{94.82\%}$&$\textbf{0.00176}$&$12.04\%$ &$0.2271$\\
$0.02$ & $94.82\%$&$0.01831$&$2.96\%$ &$0.2287$\\
$0.2$ & $94.82\%$&$0.15334$&$1.49\%$ &$0.2314$\\
$0.5$ & $94.64\%$&$0.42992$&$1.51\%$ &$0.2321$\\
$1.0$ & $56.35\%$&$0.95415$&$\textbf{1.12\%}$ &$\textbf{0.2330}$\\
\hline
\end{tabular}
\end{table}
We have considered five distinct limitations regarding the $ell_2$ norm of perturbations, which are introduced to the confidence vectors during the poisoning process. Figure~\ref{fig:magnitude} provides a visual comparison of the performance across different perturbations applied to the confidence vectors. The first row of the figure presents the original ground truth images for reference. The second row illustrates the results obtained without any defensive measures in place. Subsequent rows depict the reconstruction outcomes of the inversion model when it is subjected to the LPA defense, with each row differentiated based on the magnitude of the perturbations applied. Our findings indicate that the LPA defense is effective in safeguarding classifiers against inversion attacks and in reducing the quality of the reconstructed images. Even with a minute perturbation magnitude, such as $0.002$, the reconstructed images cannot be recognized by humans. 

Additionally, Table~\ref{tab:magnitude} presents a quantitative analysis of our findings. It is evident that an increase in the magnitude of $\delta$ correlates with a reduction in the utility of the target model and a corresponding increase in privacy protection. Specifically, as $\delta$ grows, there is a progressive enhancement in the distortion of confidence scores. Nevertheless, it is only when $\delta$ surpasses $0.2$ that we observe a decline in model accuracy. In other words, the impact of perturbation magnitude on the model accuracy is not apparent. Furthermore, as $\delta$ increases, there is a gradual decrease in attack accuracy accompanied by a steady rise in reconstruction loss. This suggests that the privacy risks associated with model inversion attacks are effectively mitigated. Overall, the perturbation $\delta$ determines the trade-off between the utility and privacy of the target model. A modest perturbation (e.g., $\delta=0.2$) can lead to robust defenses without significantly affecting model accuracy, despite the rise in confidence score distortion. 

\subsubsection{Comparison ayalysis with noise-based defenses.}
Our defense strategy augments privacy preservation by introducing perturbations to the confidence vectors. In this experiment, we delve into the comparative efficacy of various perturbation techniques. Initially, we begin our ablation study by using two methods: the commonly employed PGD adversarial attack and a poisoning attack, Poi\_MSE, which calculates the defensive gradient using the MSE loss function. Furthermore, we also compare our method with additional SOTA defenses that protect confidence vectors through noise integration. Two prominent SOTA defenses serve as benchmarks for this comparative study: AdvExa, introduced by Wen et al.~\cite{wen2021defending}, and OnePara, presented by Ye et al.~\cite{ye2022one}. A quantitative assessment is illustrated in Table~\ref{tab:moreSOTA}. The results demonstrate that a naive application of PGD can adversely impact utility, notably the model's classification precision. Furthermore, substituting our cosine similarity measure with MSE loss would reduce the defense's strength, weakening the privacy protection offered by data poisoning techniques. Moreover, Wen et al. crafted AdvExa, a sophisticated multi-step adversarial framework that embeds noise within confidence vectors, thereby retaining classification accuracy unaffected. Despite AdvExa's maintenance of accuracy, the resulting change in confidence levels falls short when compared to our proposed LPA method.

Moreover, AdvExa's privacy safeguarding capabilities are somewhat constrained, as adversarial noise is generally tailored for the specific trained inversion model. Yet, noise-based defenses are typically engineered prior to the inversion model's training phase. Thus, the consequent training procedure would significantly decrease the effectiveness of crafted noise. Ye et al.'s OnePara introduces differential privacy (DP) noise into the confidence vectors, and as depicted in Table~\ref{tab:moreSOTA}, it substantially enhances privacy measures. Nonetheless, this enhancement in privacy comes at the cost of a significant expansion in confidence distances, which may in turn limit the practical utility of the target model.

\begin{table}[!t]
\caption{Comparative Assessment of Defense Mechanisms based on Noise Injection. 
\label{tab:moreSOTA}}
\centering
\renewcommand{\arraystretch}{1.3} 
\begin{tabular}{|c|c|cc|cc|}
\hline
\multirow{2}{*}{\textbf{Dataset}}&\multirow{2}{*}{\textbf{Methods}}&\multicolumn{2}{c|}{\textbf{Utility}}&\multicolumn{2}{c|}{\textbf{Privacy}}\\
\cline{3-6}
&&\makecell[c]{Model\\Acc.}&\makecell[c]{Conf.\\Dist.}&\makecell[c]{Attack\\Acc.}&\makecell[c]{Recon.\\Error}\\
\hline
\hline
\multirow{5}{*}{MNIST}&PGD &$93.67\%$&$0.09079$&$80.05\%$&$0.88869$\\
&Poi\_MSE&$99.98\%$&$0.00014$&$79.17\%$&$0.88373$\\
\cline{2-6}
&Our &$\textbf{99.98\%}$&$\textbf{0.00014}$&$74.27\%$&$0.88915$\\
\cline{2-6}
&AdvExa&$99.98\%$&$0.05424$&$82.54\%$&$0.88817$\\
&OnePara &$99.98\%$&$0.99121$&$\textbf{20.61\%}$&$\textbf{0.89568}$\\
\hline
\hline
\multirow{5}{*}{\makecell[c]{Fashion-\\MNIST}}&PGD & $94.13\%$&$0.00115$&$47.10\%$&$0.61329$\\
&Poi\_MSE & $94.72\%$&$0.00113$&$50.64\%$&$0.61324$\\
\cline{2-6}
&Our &$\textbf{94.72\%}$&$\textbf{0.00021}$&$40.07\%$&$0.62192$\\
\cline{2-6}
&AdvExa & $94.72\%$&$0.00113$&$47.64\%$&$0.61154$\\
&OnePara &$94.72\%$&$0.97292$&$\textbf{19.96\%}$&$\textbf{0.70026}$\\
\hline
\end{tabular}
\end{table}

\begin{figure}[!h]
    \centering
    \subfloat[]{\includegraphics[width=0.435\linewidth]{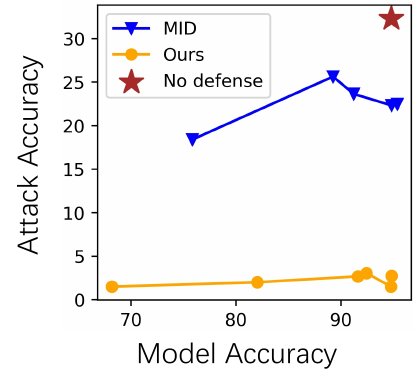}\label{fig:curve_a}}
    \subfloat[]{\includegraphics[width=0.473\linewidth]{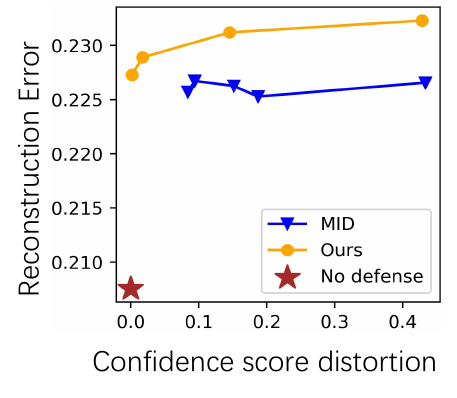}\label{fig:curve_b}}
    \caption{The trade-off between utility and privacy under different defenses.}
    \label{fig:curve}
\end{figure}
\subsubsection{Trade-off comparisons with MID}
Continuing to test our defense method LPA, our next experiments compared the privacy-utility trade-off of this defense to the current state-of-the-art defense, MID~\cite{MID2020}. The results on FaceScrub are provided in Fig.~\ref{fig:curve}. We find that our LPA method can outperform MID by a large margin to defend against the model inversion attacks. In particular, in Fig.~\ref{fig:curve_a}, lower attack accuracy implies higher privacy protection, whereas higher model accuracy denotes higher utility. The attack accuracy under our LPA defense is consistently much lower than it is under MID. Our defense poses little impact on the model accuracy, and the attack accuracy can be decreased to a low level even though the model accuracy is high. Therefore, when compared to MID, LPA effectively improve the trade-off between attack accuracy and model accuracy. Additionally, Fig.~\ref{fig:curve_b} shows that our LPA can also outperform MID in terms of reconstruction error for a given level of confidence score distortion, where higher reconstruction error suggests lower privacy risks and lower confidence score distortion implies higher utility. Note that the LPA method does not require retraining and is hence more applicable. Consequently, LPA may be preferable due to its superior trade-off and flexibility.

\subsubsection{Time efficiency of LPA}
\textcolor{black}{The time efficiency of defensive methods is crucial for their practical application. We conducted experiments to assess the time overhead introduced by our LPA defense. The experimental results are presented in Table~\ref{tab:time}. The second column, `Training', indicates the time required to train the classifier, measured in seconds. The `Inference' column shows the time needed to make a prediction for each sample during the inference stage, measured in milliseconds. Additionally, `Poisoning' represents the time overhead introduced by the defensive measure, specifically the perturbation calculation, in LPA for each sample.}

\textcolor{black}{Our results demonstrate that while the LPA method introduces some additional latency, its time overhead remains acceptable, especially considering it eliminates the need for costly victim model retraining. For datasets such as MNIST, FashionMNIST, and CIFAR-10, the defensive measure (i.e., perturbation calculation) in LPA introduces a latency of approximately $3.6$ milliseconds per sample. Although this perturbation process significantly increases response time compared to the inference time alone, the total inference time with defensive measures remains within the millisecond range. Consequently, such minor delays are typically imperceptible in real-world applications, ensuring a seamless user experience. For the FaceScrub dataset, the total inference time also remains in the millisecond range after deploying the LPA defense.}
\begin{table}[!t]
\caption{Time Efficiency Assessment of the LPA Defense. We report the time during classifier training, inference, and defense stage in the format of ``mean(standard deviation)". 
\label{tab:time}}
\centering
\renewcommand{\arraystretch}{1.5} 
\begin{tabular}{|c|c|c|c|}
\hline
\textbf{Datasets} & \textbf{Training} & \textbf{Inference} & \textbf{Poisoning} \\
\hline
MNIST & $387.2$s $(4.03)$ & $0.35$ms $(0.02)$ & $3.69$ms $(0.18)$\\
FashionMNIST & $390.4$s $(2.49)$ & $0.39$ms $(0.02)$ & $3.61$ms $(0.16)$\\
CIFAR10 & $398.1$s $(2.29)$ & $0.36$ms $(0.02)$ & $3.59$ms $(0.12)$\\
FaceScrub & $679.7$s $(3.28)$ & $0.38$ms $(0.01)$ & $9.71$ms $(0.85)$\\
\hline
\end{tabular}
\end{table}

\begin{figure}[!h]
    \centering
    \includegraphics[width=0.7\linewidth]{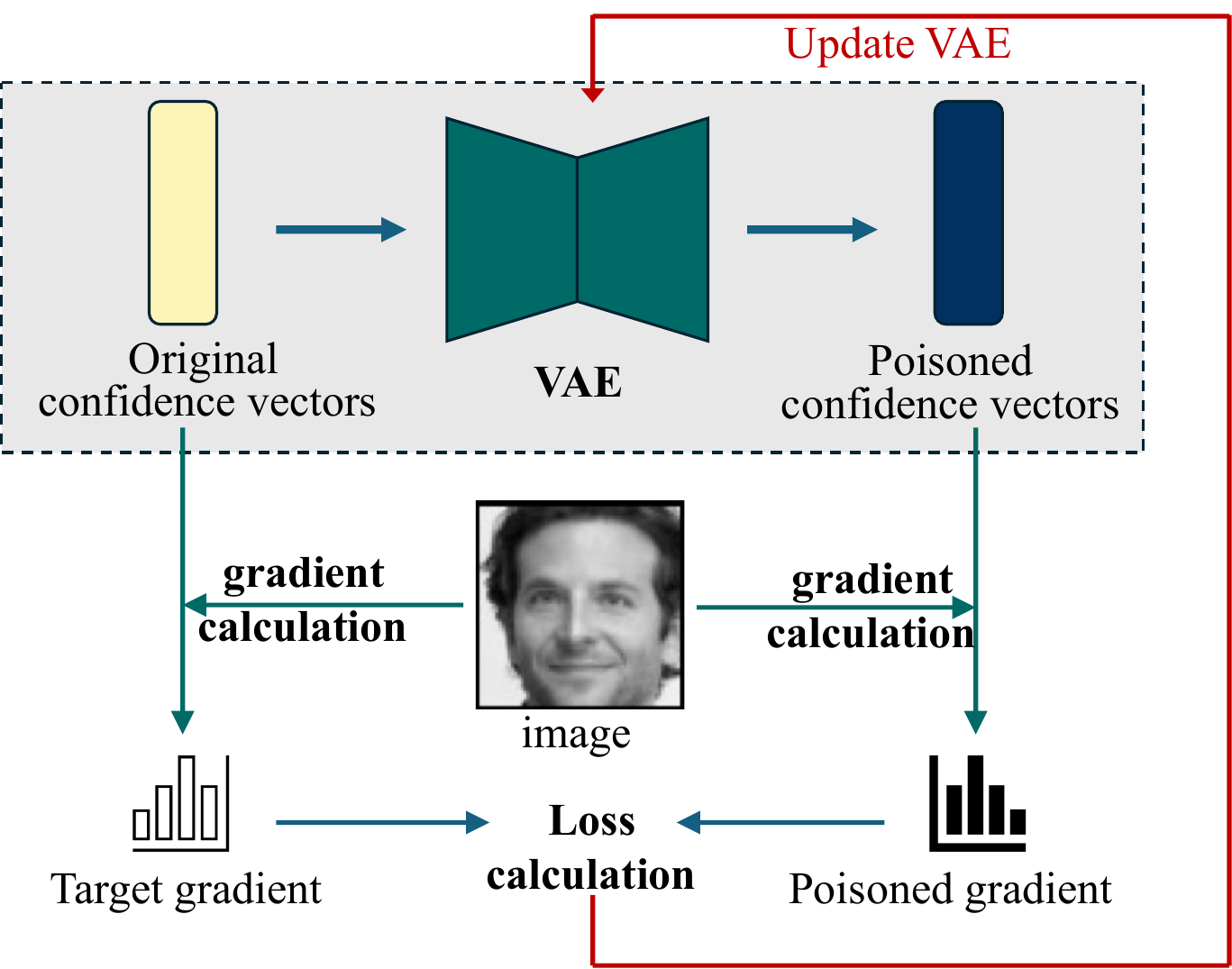}
    \caption{Replace the optimization process via a VAE to improve the time efficiency of defenses.}
    \label{fig:VAE}
\end{figure}

\textcolor{black}{To further optimize time overhead in time-sensitive applications, we propose an enhancement using a Variational Autoencoder (VAE). The VAE is trained to capture and reproduce the distribution of defensive perturbations that would normally be generated through optimization process in LPA. Once trained, this VAE enables rapid perturbation generation through a single forward pass, eliminating the need for iterative optimization. The training process of this VAE is illustrated in Fig.~\ref{fig:VAE}. The VAE is designed to transform original confidence vectors into their corresponding poisoned variants. To guide the VAE training, we employ a specialized loss function based on the negative cosine similarity between two gradient vectors: the target gradient and the poisoned gradient. The target gradient is computed using the original confidence vector paired with its corresponding input image, while the poisoned gradient is derived from the VAE-generated poisoned confidence vector and the same input image. Through iterative training, the VAE learns to generate poisoned confidence vectors containing the necessary protective perturbations.} 

\textcolor{black}{To validate the effectiveness of this VAE-based perturbation generation approach, we conducted experimental evaluations on MNIST, with results presented in Table~\ref{tab:vae_time}. Compared to previous optimization-based perturbation generation, our VAE approach demonstrates remarkable computational efficiency, achieving a per-sample latency of only $0.34$ milliseconds while introducing a marginal increase in confidence distance metrics. This VAE-based variant is particularly suitable for applications requiring rapid response times.}

\begin{table}[!t]
\caption{Time Efficiency Assessment of the VAE-based generation of perturbations. The baseline scenario without defenses is denoted as `No-def.' and `Opti.' means that perturbations are derived through iterative optimization procedures.
\label{tab:vae_time}}
\centering
\renewcommand{\arraystretch}{1.3} 
\begin{tabular}{|c|cc|cc|c|}
\hline
\multirow{2}{*}{\textbf{Methods}}&\multicolumn{2}{c|}{\textbf{Utility}}&\multicolumn{2}{c|}{\textbf{Privacy}}&\multirow{2}{*}{\textbf{Time}}\\
\cline{2-5}
&\makecell[c]{Model\\Acc.}&\makecell[c]{Conf.\\Dist.}&\makecell[c]{Attack\\Acc.}&\makecell[c]{Recon.\\Error}&\\
\hline
\hline
No-def. &$99.98\%$&$-$&$80.05\%$&$0.88357$& $-$ \\
\hline
Opti. &$99.98\%$&$\textbf{0.00014}$&$\textbf{79.17}\%$&$\textbf{0.88915}$& $3.69~(0.18)$\\
VAE &$99.98\%$&$0.00196$&$79.85\%$&$0.88909$& $\textbf{0.34}~(0.05)$ \\
\hline
\end{tabular}
\end{table}

\subsubsection{Defensive effectiveness}
\begin{figure}[!t]
    \centering
    \includegraphics[width=0.95\linewidth]{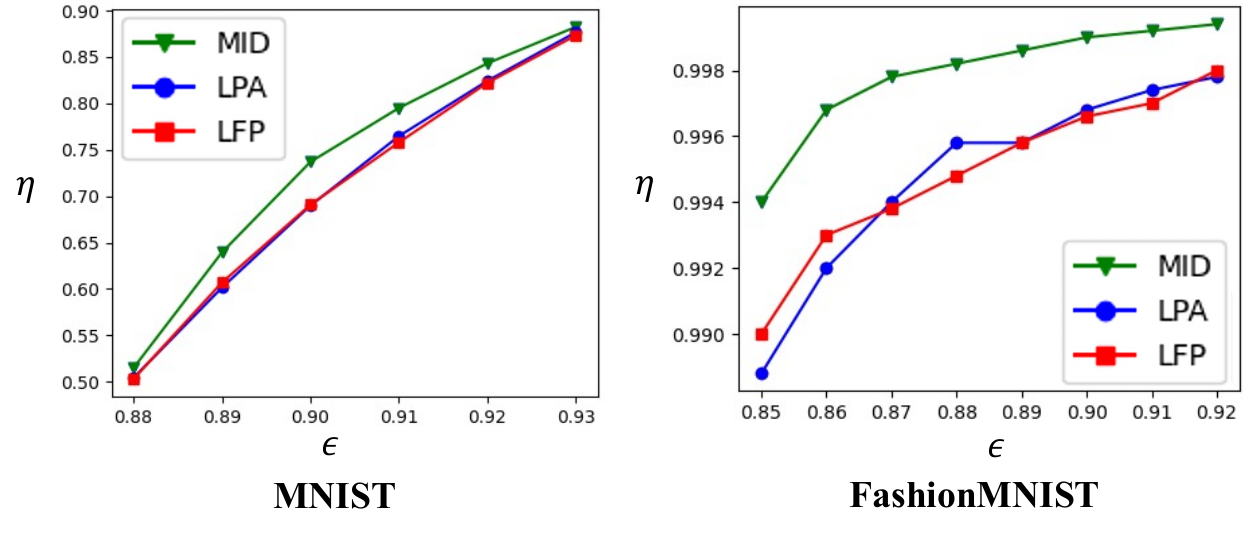}
    \caption{The defensive effectiveness from the perspective of $\epsilon-\eta$ defense on three datasets.}
    \label{fig:eta}
\end{figure}

We also conducted simulation experiments based on our proposed definition, $\epsilon-\eta$ defense to assess the effectiveness of different defenses. To explore the relationship between $\epsilon$ and $\eta$ on three datasets, we calculated the fraction of images with reconstruction errors exceeding $\epsilon$, representing defense success rate (i.e., $1-\eta$). Intuitively, $\epsilon$ can be thought of as a threshold, where a greater $\epsilon$ results in fewer reconstructed data with a reconstruction error greater than the threshold. Hence, $\epsilon$ should be negatively proportional to the probability $1-\eta$. Our experiments validated this intuition. As shown in Fig.~\ref{fig:eta}, $\eta$ increases as the reconstruction error threshold $\epsilon$ improves.

Additionally, for a fixed reconstruction error threshold $\epsilon$, a lower $\eta$ signifies a higher proportion of reconstructed samples exceeding this threshold, indicating superior defensive performance. As illustrated in Fig.~\ref{fig:eta}, MID method typically results in a greater $\eta$ when compared to our LPA and LFP defenses. This experimental result shows that the performance of LPA and LFP are comparable and both of them outperform MID method according to this $\epsilon-\eta$ defense metric.

\subsubsection{Robustness against adaptive attacks}
\textcolor{black}{We evaluated the robustness of LPA against adaptive attackers who were aware of our LPA defense. These attackers employed two methods to transform the confidence vectors in attempts to destroy the poisoning ability of the LPA defense. The first method introduced random noise into the LPA-processed confidence vectors, while the second method applied a SoftMax function to produce new confidence vectors.}

\textcolor{black}{As shown in Table~\ref{tab:adaptive}, our experimental results demonstrate that LPA remains robust against both types of adaptive attacks. Rather than compromising the poisoning ability, these transformation methods disrupted the underlying mapping between private image data and confidence vectors. Consequently, the inversion model's training was compromised, resulting in reconstructed images with lower attack accuracy and higher reconstruction error compared to the baseline LPA defense. These findings align with previous research that employed differential privacy mechanisms to protect against privacy leakage by perturbing the image-confidence mapping~\cite{ye2022one}. In summary, our results confirm that LPA effectively maintains its defensive capabilities against adaptive attacks.}
\begin{table}[!t]
\caption{Robustness evaluation against two types of adaptive attacks, including introducing random noise (denoted as `random') and applying the SoftMax function (denoted as `softmax').
\label{tab:adaptive}}
\centering
\renewcommand{\arraystretch}{1.3}
\begin{tabular}{|c|c|cc|cc|}
\hline
\multirow{2}{*}{\textbf{Dataset}}&\multirow{2}{*}{\textbf{Methods}}&\multicolumn{2}{c|}{\textbf{Utility}}&\multicolumn{2}{c|}{\textbf{Privacy}}\\
\cline{3-6}
&&\makecell[c]{Model\\Acc.}&\makecell[c]{Conf.\\Dist.}&\makecell[c]{Attack\\Acc.}&\makecell[c]{Recon.\\Error}\\
\hline
\hline
\multirow{3}{*}{MNIST}&LPA &$99.98\%$&$\textbf{0.00014}$&$74.72\%$&$0.88915$\\
&random &$99.98\%$&$0.00021$&$44.83\%$&$0.89152$\\
&softmax &$99.98\%$&$0.66523$&$\textbf{41.51\%}$&$\textbf{0.89614}$\\
\hline
\multirow{3}{*}{FaceScrub}&LPA &$94.82\%$&$\textbf{0.00340}$&$5.08\%$ &$0.23181$\\
&random & $94.82\%$&$0.00343$&$4.96\%$&$0.23372$\\
&softmax & $94.82\%$&$0.90809$&$\textbf{2.43\%}$ &$\textbf{0.23921}$\\
\hline
\end{tabular}
\end{table}

\bibliographystyle{IEEEtran}
\bibliography{main}

\end{document}